\documentclass[%
 reprint,
 amsmath,amssymb,
 aps,prl
]{revtex4-1}

\renewcommand\section[1]{}

\usepackage{xcolor}

\usepackage{soul}
\usepackage{cancel}

\usepackage{xr}

\usepackage{graphicx}
\usepackage{dcolumn}
\usepackage{bm}

\begin{document}

\preprint{APS/123-QED}

\title{Universal and accessible entropy estimation using a compression algorithm}

\author{Ram Avinery}
\email{ramavine@mail.tau.ac.il}
\author{Micha Kornreich} 
\altaffiliation[Currently at ]{Department of Physics, New York University, New York, New York 10003, USA.}
\author{Roy Beck}
\email{roy@tauex.tau.ac.il}
\affiliation{The Raymond and Beverly Sackler School of Physics and Astronomy, Tel Aviv University, Tel Aviv 69978, Israel}%


\begin{abstract}
Entropy and free-energy estimation are key in thermodynamic characterization of simulated systems ranging from spin models through polymers, colloids, protein structure, and drug-design. Current techniques suffer from being model specific, requiring abundant computation resources and simulation at conditions far from the studied realization. Here, we present a universal scheme to calculate entropy using lossless compression algorithms and validate it on simulated systems of increasing complexity. Our results show accurate entropy values compared to benchmark calculations while being computationally effective. In molecular-dynamics simulations of protein folding, we exhibit unmatched detection capability of the folded states by measuring previously undetectable entropy fluctuations along the simulation timeline. Such entropy evaluation opens a new window onto the dynamics of complex systems and allows efficient free-energy calculations.
\end{abstract}

\maketitle


\section{\label{sec:level1I}INTRODUCTION}

Utilizing the exponentially growing power of computers enables \textit{in-silico} experiments of complex and dynamic systems \cite{Dror2012}. In these systems, entropy ($S$) and enthalpy ($H$) should be evaluated to appraise the system thermodynamic properties. While enthalpy can be directly calculated from the interaction strength between the system's components, computing the entropy of an equilibrated canonical system essentially requires inferring the probabilities of all relevant microstates (\textit{i.e.}, specific configurations). Consequently, for large systems, contemporary computational capabilities struggle to simulate sufficient microstates for adequate mapping of the free-energy landscape. This fact limits current ability to estimate thermodynamic properties of interesting systems and phenomena including, \textit{e.g.}, protein folding \cite{Dror2012,Piana2012,Piana2014}.
 
Present strategies to estimate the entropy from simulations include density- or work- based methods \cite{Kofke2005}. These methods have been proven useful, though they rely on plentiful computational power, for simulations away from the designated realization \cite{landau2014guide,Frenkel2001,Kubelka2006}. Notably, no single method for entropy and free-energy evaluation can be viewed as superior to others, and in many cases, the choice is system dependent \cite{Hansen2014}. As an alternative path, a reduced phase space assignment can be used, as previously demonstrated in protein folding simulations \cite{Dror2012,Piana2012,Buchete2008}. There, using \textit{a priori} knowledge, such as the experimental native protein structure, can be used to attribute each frame a specific state (\textit{e.g.}, folded or unfolded protein states). Following, a rough estimate of the system's free-energy and thermodynamic properties is then attained using the respective state populations, although entropy values are not directly assigned. 

Seminal papers in information theory by Shannon \cite{Shannon1948} and Kolmogorov \cite{Kolmogorov1968} introduced measures of uncertainty which are mathematically identical to the statistical-mechanics definition of entropy at the large dataset limit. Lossless compession algorithms are essentially practical implementations attempting to realize Kolmogorov complexity \cite{downarowicz2011entropy,krieger1970entropy,krieger1970entropy}.
Recognizing these relations has produced novel analytical methods for studying mutual information in sequences of symbols, internet traffic analysis, and redundancy anomaly detections for medicinal signal analysis in electroencephalography,  electrocardiography, and more \cite{Henriques2013,Aboy2006,Benedetto2002,Amigo2004}. Despite these important links, studies of physical systems using lossless-compression are rather sparse. Exceptions include recent studies on thermodynamic phase transitions \cite{Melchert2015,Vogel2012,Vogel2017,Martiniani2019}. Additional details on previous studies involving compression algorithms for physical systems are given in the Supplementary Material \cite{supp}.

Here, we present \textit{a framework for accessible and accurate asymptotic entropy ($S_\mathrm{A}$) calculation using a lossless compression algorithm}. Conceptually, the redundancy of information stored in a recorded simulation is tightly related to the entropy of the physical system being simulated. At the foundation of our method, we use a lossless compression algorithm which is optimized to remove information redundancy by locating repeated patterns within a stream of data. Thus, the ability to compress a digital representation of a physical system is directly related to the entropy of that system \cite{supp, CoverTM2006EOIT}. We note that other methods exist for the estimation of information-entropy in a data stream \cite{CoverTM2006EOIT}. Here we chose to utilize lossless compression, due to its availability and ease of use. 
 
As a proof-of-concept, we verify our entropy estimation on various model systems where the entropy is analytically calculated and compared. Later, the direct application of asymptotic entropy calculation is demonstrated on protein folding simulations, where entropy estimation is challenging.

Most adopted lossless compression implementations derive from schemes introduced by Lempel and Ziv (LZ) \cite{Ziv1977,Ziv1978,pu2005fundamental}. LZ algorithms process an input sequence of symbols in a finite alphabet and produce a compressed output sequence by replacing short segments with a reference to a previous instance of the same segment. Fundamentally, the ratio of LZ compressed to input sequence lengths has been proven to converge to Shannon's entropy definition \cite{Lesne2009,Benedetto2002}. This convergence is guaranteed for an infinite sequence of symbols, produced by an ergodic random source. A sequence of independent microstates sampled from a physical system in equilibrium is in accord with the required random source \cite{supp}.

One expects LZ schemes to produce an upper bound on physical entropy and approach it asymptotically for large datasets \cite{Lesne2009}. In practice, our entropy estimation converges to within a few percents from expected values, even for relatively small datasets. This result, in combination with readily available enthalpy values from the simulations, allows us to construct enthalpy-entropy population diagrams for the complex and dynamic simulation of protein folding.

\section{\label{sec:level2}METHODOLOGY AND RESULTS}

To calculate entropy using a compression-based algorithm, we must quantitatively map the information content (compressed length) to entropy in the proper scale. However, preliminary steps are required to eliminate spurious effects that result from the combination of translating physical systems into 1d datasets, the physical nature of the specific problem, and the algorithm limitations.

Several physical systems are represented using continuous variables. There, each variable requires an enormous alphabet to represent each degree of freedom. This poses a difficulty for compression since the least-significant digits are noisy, hence incompressible. Therefore, a preprocess is required to reduce the alphabet variability to a coarse-grained representation with $n_\mathrm{s}$ values. For additional preprocessing details see \cite{supp}.

Following, we now take the discretized configurations and store them contiguously in a 1d file \cite{supp}. We define the original and compressed file sizes, measured in bytes, by $\tilde{C_\mathrm{d}}$ and $C_\mathrm{d}$ respectively. To properly evaluate the asymptotic entropy $S_\mathrm{A}$, we generate two additional datasets having the original dataset length. In the first, data over the entire phase-space is replaced with a single repeating symbol (\textit{e.g.}, zero). In the second, all the dataset is replaced with random symbols from the alphabet. The resulting two compressed dataset file sizes are denoted by $C_0$ and $C_1$, respectively. The ratios  $C_0/\tilde{C_0}$ and $C_1/\tilde{C_1}$ converge at the large dataset limit to a value that depends on the size of the alphabet \cite{supp}.

Since the degenerate and random datasets represent the extreme cases of minimal and maximal entropy, the compressed file size for the simulated state ($C_\mathrm{d}$) lays within these two extremes. Therefore, we define the incompressibility by $\eta=(C_\mathrm{d}-C_0)/(C_1-C_0 )$. For physical systems  $0\leq\eta\leq1$, and converges to a constant in equilibrium with sufficient sampling.  

\begin{figure}
	\includegraphics[width=0.5\textwidth]{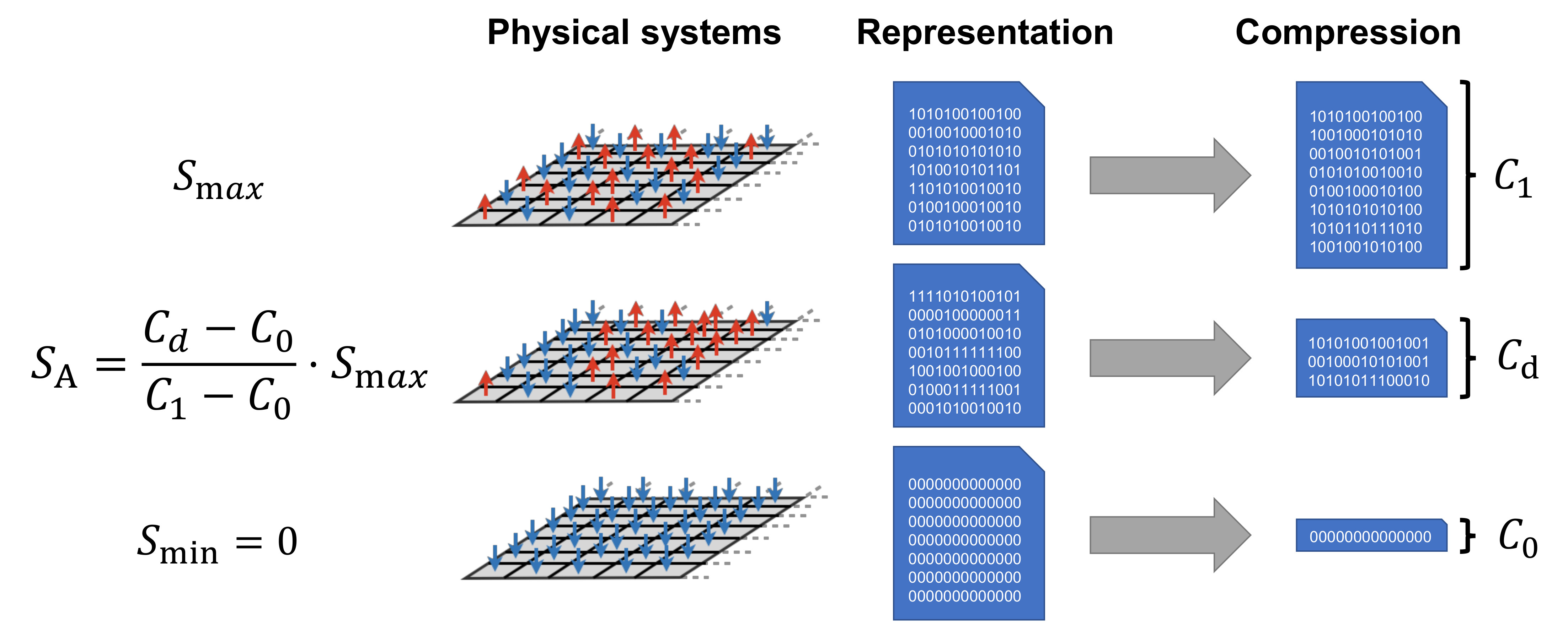}
	\caption{\label{fig:scheme} Schematic asymptotic entropy calculation. Simulations of physical systems are preprocessed and encoded into data files \cite{supp}. Entropy is directly calculated from the size of the compressed ($C_\mathrm{d}$) and calibration ($C_0, C_1$) data, as well as the entropy range ($S_\mathrm{min},S_\mathrm{max}$).}
\end{figure}

Finally, mapping $\eta$ to $S_\mathrm{A}$ can be conducted in various ways, for example from prerequisite knowledge on specific entropy values. Alternatively, we recognize that for each of the $D$ degrees of freedom in the system, represented with $n_\mathrm{s}$ discrete values, the maximal entropy is given by $k_\mathrm{B}\log n_\mathrm{s}$, where $k_\mathrm{B}$ is the Boltzmann constant. Therefore, as a first order approximation, we linearly map $\eta$ to entropy, up to an additive constant, by taking $S_\mathrm{A}/k_\mathrm{B}=\eta D\log n_\mathrm{s}$ (Figure \ref{fig:scheme}) \cite{supp}. Below we demonstrate that this linear mapping asymptotically quantifies the entropy even with finite sampling and far from the large dataset limit (e.g., number of microstates). 

\begin{figure}
	\includegraphics[width=0.5\textwidth]{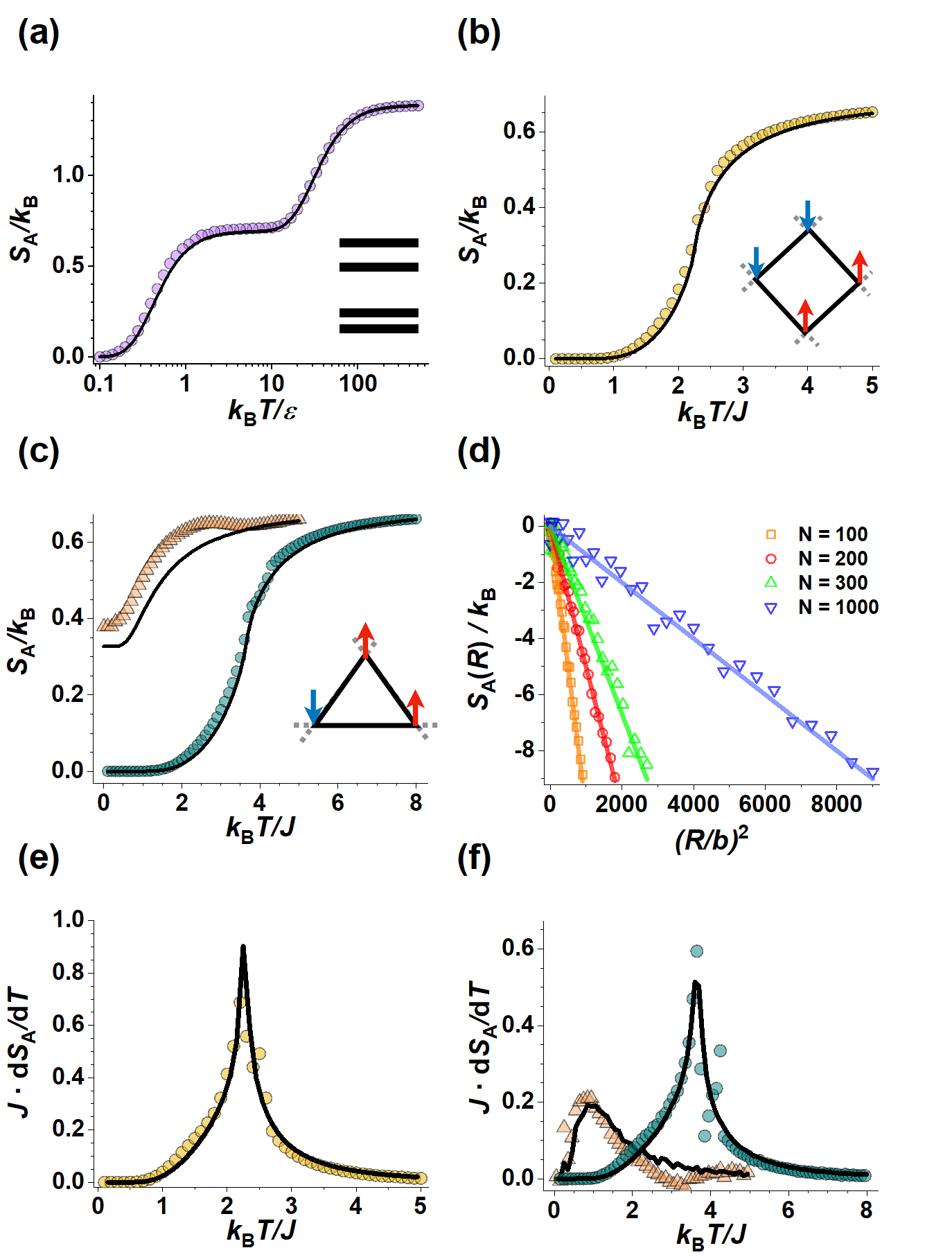}
	\caption{\label{fig:one} Validation to asymptotic entropy calculation to Monte-Carlo simulation data on benchmark model systems, by comparing analytical entropy calculation (lines) and compression algorithm method (symbols). (a) Discrete energy levels; Ising models on (b) square lattice, and (c) triangular lattice, either with antiferromagnetic ($J<0$, triangles) or ferromagnetic ($J>0$, circles); (d) Ideal chains held at varying end-to-end distance ($R$). (e) and (f) Entropy derivatives of the Ising model simulations [(b) and (c) respectively].}
\end{figure}

\begin{figure}[b]
	\includegraphics[width=0.5\textwidth]{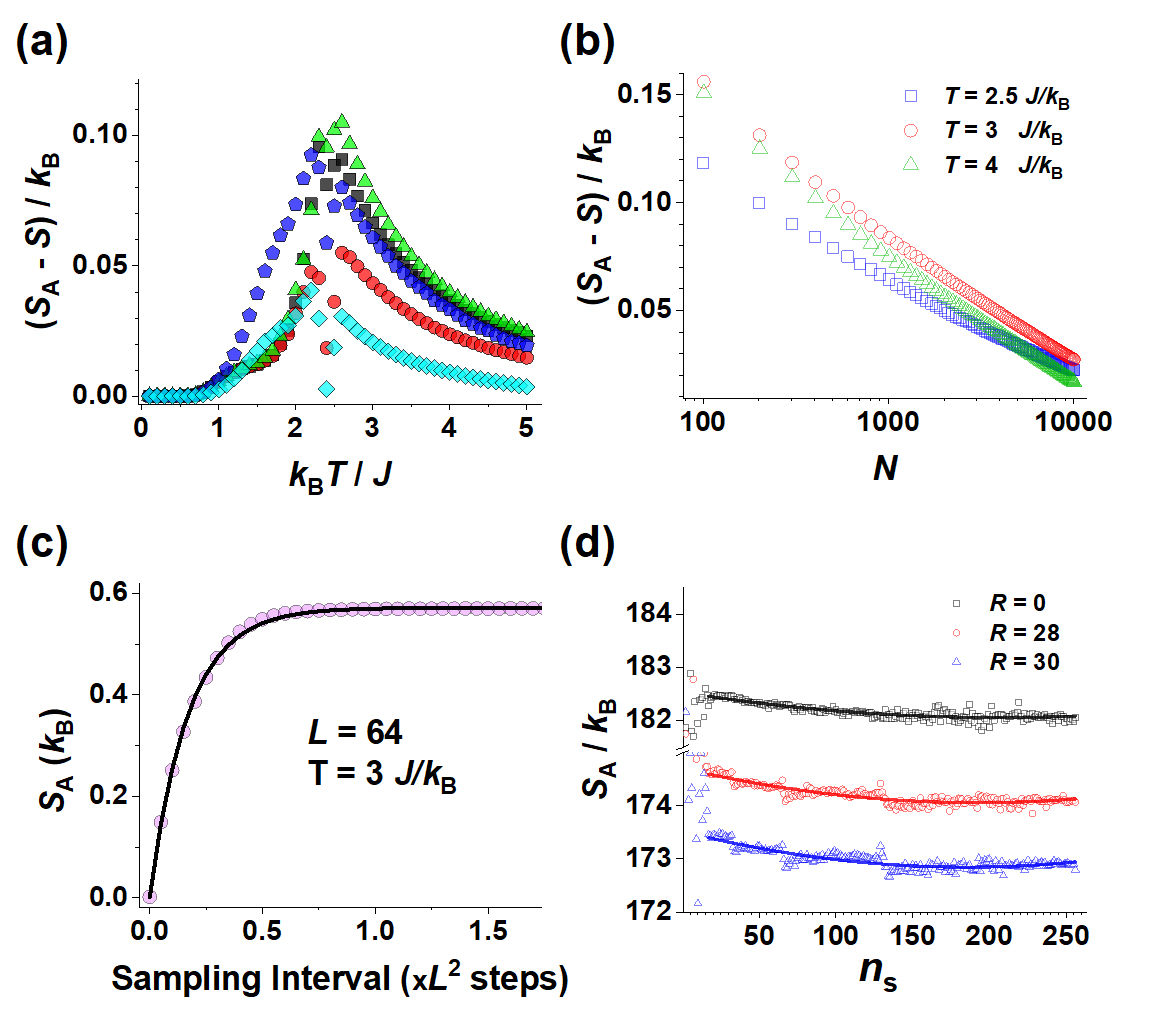}
	\caption{\label{fig:two} (a) Residuals from analytical entropy for various 1d reductions of the Ising model on a square lattice: row by row (squares), spiral (triangles), Hilbert (circles), Hilbert with 1 site/byte (pentagons), Hilbert with 3 sites/byte (diamonds). (b) and (c) Asymptotic entropy convergence for the Ising model on a square lattice. (b) Varying sample size ($N$) and fixed sampling interval ($140 L^2$). (c) Varying sampling interval with fixed sample size ($10^{4}$), exhibiting exponential convergence (solid line). (d) Effect of coarse-graining level on $S_\mathrm{A}$  for the ideal chain. Lines are a guide for the eye.}
\end{figure}

We are now ready to evaluate our scheme for several benchmark systems. Herein, we use the LZMA compression algorithm although other algorithms produce qualitatively similar results [19]. We compare $S_\mathrm{A}$ to analytical entropy calculation of five different systems [Figs. \ref{fig:one}(a-f)]: finite energy levels ($\epsilon,2\epsilon,70\epsilon,80\epsilon$) with an arbitrary energy scale ($\epsilon$) simulated at different temperatures ($T$), a 2d Ising model on a square lattice, 2d ferromagnetic and antiferromagnetic (frustrated) Ising models on triangular lattices, and an ideal chain fluctuating in 2d with fixed end-to-end distance ($R$). The Ising models have exchange energy $J$, the ideal chain is simulated with monomer length $b$, and all systems are simulated using Monte-Carlo algorithms \cite{supp}. The results agree well with the theoretical calculation [Figs.  \ref{fig:one}(a-d)]. In fact, for the Ising model on a square lattice, maximal residues from analytical values are smaller than $0.04 k_\mathrm{B}$ [Fig.  \ref{fig:two}(a)]. For the ideal chain simulation, our entropy estimation matches the known entropy dependence of $S(R)-S(0)=-R^2/b^2 (N-1)$, where $N$ is the number of monomers, without any fitting parameters [Fig. \ref{fig:two}(e)] \cite{supp}. Also, our results present a smooth trend and enable to differentiate $S_\mathrm{A}$ for specific heat and critical exponent derivations [Figs.  \ref{fig:one}(e-f)].

Since compression algorithms result in an upper bound for the entropy, we can evaluate and optimize different preprocessing protocols \cite{supp}. For example, a comparison between different 2d to 1d transformations for the Ising model on a square lattice shows that the Hilbert scan \cite{Moon2001} is slightly better than other naive transformations [Fig.  \ref{fig:two}(a)]. Notably, we can use data compression to evaluate ergodicity and proper sampling intervals [Fig.  \ref{fig:two}(b)] \cite{Lesne2009, supp}. While the convergence of $S_\mathrm{A}$ with increasing sampling interval is exponential, its convergence with additional sampling is logarithmic [Fig.  \ref{fig:two}(c)], as expected \cite{Plotnik1992UpperBounds}, and will level off as it approaches the actual value, similarly to trends in random data Fig. \ref{fig:five} \cite{supp}. Moreover, for as low as 1000 frames $S_\mathrm{A}$ estimate is a few percents off the analytically calculated values. 

Next, we consider the case of continuous variables which must be coarse-grained for further processing and apply it on simulated lattice-free ideal chains \cite{supp}. The optimal discretization ($n_\mathrm{s}$) should depend on the correlations in the system and the number of sampled configurations. Furthermore, the choice of a coordinate system representing the degrees of freedom in the system can introduce or eliminate correlations. In our case, the ideal chain simulation is recorded with 64-bit floating numbers for each Cartesian coordinate, but the analysis is applied to the 1d bond-angle representation \cite{supp}. We note that standard compression algorithms work best with short range correlations. For physical systems exhibiting long-range correlations, more attention will be required, possibly by transformation to an alternative representation (\textit{e.g.} Fourier transform).  

The ideal chain example validates that optimal coarse-graining can be identified using our procedure [Fig.  \ref{fig:two}(d)]. At the low $n_\mathrm{s}$ limit significant information is lost, and the entropy estimate cannot be resolved well. On the other hand, pattern-matching by the compression algorithm is hindered by finite sampling and the estimate increases towards the maximal entropy at the high $n_\mathrm{s}$ limit. Indeed, $S_\mathrm{A}$ evaluation shows a shallow minimum that deepens as the chain is stretched \cite{supp}.

Encouraged by our results, we test our entropy estimation scheme where free-energy evaluation is a serious concern, namely in protein folding simulation. There, entropy evaluation is currently limited when using the simulation data alone \cite{Singh2010}. Specifically, we quantify entropy for the reversible protein folding of a Villin headpiece C-terminal fragment simulated by molecular dynamics (MD) \cite{Piana2012}. The system is sampled at equilibrium and demonstrates short transition times between folded and unfolded states and a long lifetime at each given state \cite{Piana2012}. Piana \textit{et al.} \cite{Piana2012} calculated the fragment's thermodynamic properties from the population ratio of folded to unfolded states via the “transition-based assignment” \cite{Buchete2008} aided by the experimental folded structure \cite{Chiu2005}. In particular, the difference in entropy between folded and unfolded states ($\Delta S_\mathrm{f}$) was estimated from the states' lifetimes (Table~\ref{tab:table1}).  Using our compression framework, along with the above-mentioned frame assignments into two ensembles, we attained the backbone's entropy values of the folded and unfolded states \cite{supp}.

\begin{figure}
	\includegraphics[width=0.5\textwidth]{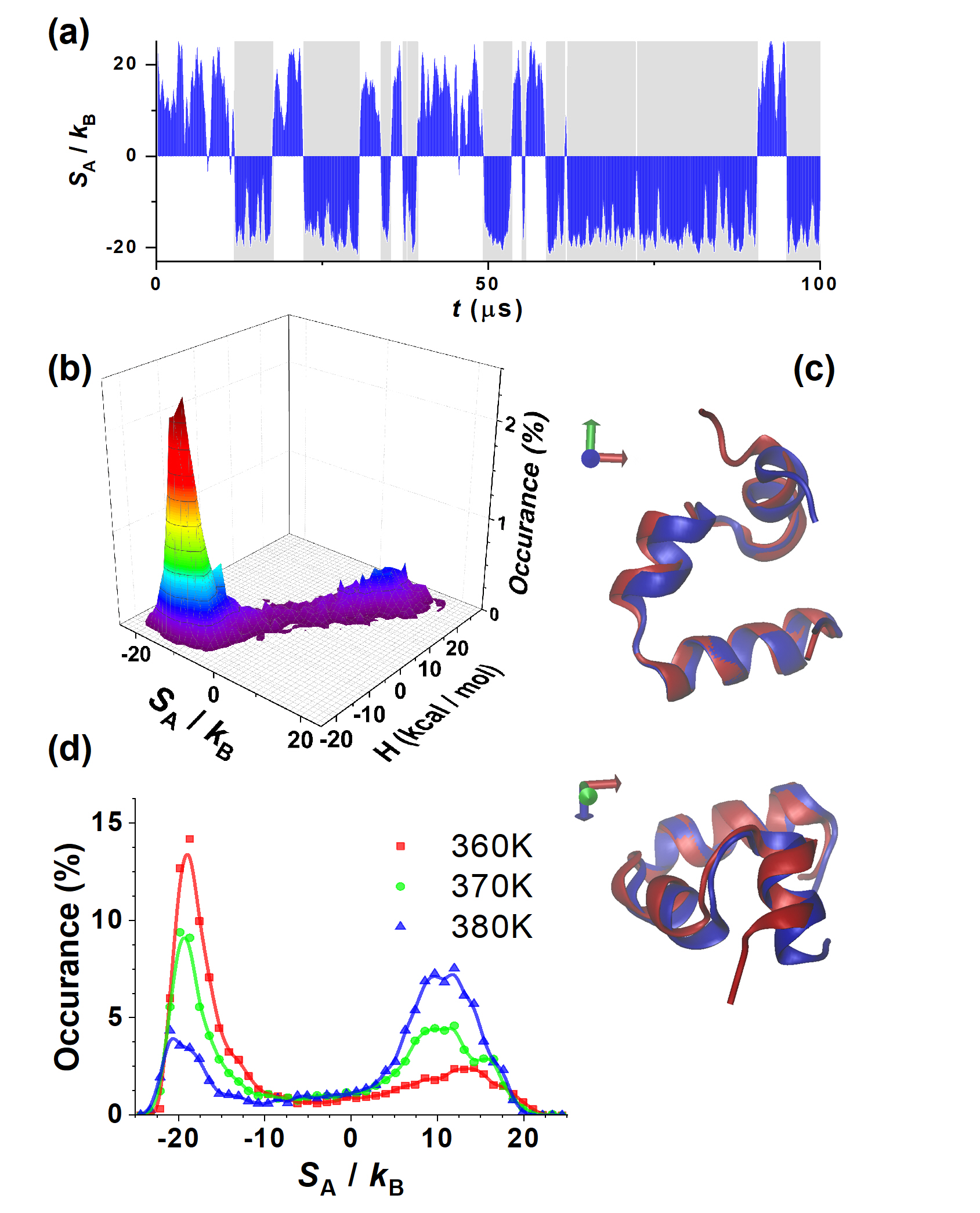}
	\caption{\label{fig:protein}(a) Representative scan through the Villin headpiece simulation timeline, with mean-subtracted $S_\mathrm{A}$ (blue bars), overlaying regions identified as folded using transition-based assignment (gray area). (b) Enthalpy -- entropy population diagram at $T$ = 360 K. (c) Protein structure collected from randomly sampled low $S_\mathrm{A}$ states simulated at 360 K (red) overlaid with the protein fragment (2F4K) crystal structure (blue) \cite{Kubelka2006,Chiu2005}, generated with PyMOL \cite{PyMOL}. (d) Distributions of sliding-window $S_\mathrm{A}$, at three simulation temperatures.}
\end{figure}

Moreover, the assignment to either of the two states can be done using a sliding entropy estimate from lossless compression, without any \textit{a-priori} experimental input. This scheme can potentially detect yet unidentified, competing, low free-energy structures which eluded experimental observation. In Fig. \ref{fig:protein}(a) we show $S_\mathrm{A}$ evaluated for sequences of configurations within a sliding window of length $\tau_\mathrm{w}$ through the timeline of a simulation. At each time point $t$, configurations sampled by the simulation between $t$ and $t + \tau_\mathrm{w}$ are preprocessed and compressed, to arrive at $S_\mathrm{A}(t)$ \cite{supp}.
We chose the window length $\tau_\mathrm{w}=0.4 \mathrm{\mu s}$ as a reasonable compromise between convergence of $S_\mathrm{A}$ and time-resolution \cite{supp}. This choice limits our observations to dynamic processes slower than the chosen time-window. Fig. \ref{fig:protein}(a) clearly demonstrates the correspondence between low $S_\mathrm{A}$ and Piana's preassigned folded states (shaded areas).

Using these sliding-window $S_\mathrm{A}$ values, we can now construct an enthalpy-entropy population diagram [Fig. \ref{fig:protein}(b)] \cite{supp}; a valley between clustered events in the $S_\mathrm{A}-H$ plot allows to assign the folded and unfolded states, without \textit{a priori} experimental knowledge, with 95.3 -- 96.3\% agreement \cite{supp} allowing the low free-energy folded structure to be extracted from the simulation and compared to the experimental crystal structure [Fig. \ref{fig:protein}(c)]. In agreement with Piana's assignments, changes in the ratio between folded and unfolded populations are clearly revealed by $S_\mathrm{A}$ distributions, as simulation temperature is varied [Fig. \ref{fig:protein}(d), Table \ref{tab:table1}].

Following an assignment to the two folded/unfolded ensembles, we can now use our method to directly estimate the entropic difference between the ensembles.  In order to reduce spurious effects resulting from time correlations between neighboring frames, we resampled each ensemble (folded/unfolded) separately \cite{supp}. Following, we optimize the number of coarse-grained dihedral angles, as described above, and estimate each ensemble's entropy value using lossless compression. The difference between the estimated values ($\Delta S_\mathrm{A}$) is given in Table \ref{tab:table1}. We note that $\Delta S_\mathrm{A}$ represents only the protein backbone's entropic contribution, as the information is taken solely from the dihedral angles. Further contributions, originating from the solvent, side-chains or other solutes may contribute as well to the overall entropy difference. These contributions will be addressed in future work.

\section{\label{sec:level3}CONCLUDING REMARKS}
By construction, the successful operation of compression algorithms is derived from identifying domains that repeat within 1d datasets. This is of great convenience for effectively 1d objects such as polymers and proteins. We show that lossless compression algorithms allow efficient estimation of entropy in a wide variety of physical systems, including protein folding simulations, and without any a priori knowledge about specific states. Additionally, our framework can easily assess sufficient sampling, ergodicity and coarse-graining optimality for many-body simulations. We expect that our methodology will be useful for experimental systems \cite{Martiniani2019} and additional athermal models, where entropy estimation is hard or inaccessible.

Entropy is defined for equilibrated or almost-stationary systems. However, $S_\mathrm{A}$ estimates can be useful also away from equilibrium, to detect divergent trends in information-content and disorder \cite{Martiniani2019}. Our results demonstrate modern MD simulations have sufficient statistics to allow entropy estimation even for small fragments of simulated trajectories, and that lossless compression algorithms can be conveniently used for this estimation. The resulting observation of continuous entropy dynamics, including detection of transient ordered states (\textit{i.e.}, a protein's fold), opens a new avenue in characterizing dynamics of complex systems.

\begin{acknowledgments}
We greatly appreciate fruitful discussions and comments from A. Aharony, D. Andelman, P. Chaikin, H. Diamant, E. Eisenberg, O. Farago, D. Frenkel, M. Goldstein, G. Jacoby, D. Levin, R. Lifshitz, Y. Messica, H. Orland, P. Pincus, Y. Roichman, Y. Shokef, and H. Suchowski. Special thanks for David Shaw laboratory for sharing the MD data. The work is supported by the Israel Science Foundation (453/17, 550/15) and United States - Israel Binational Science Foundation (201696).

\end{acknowledgments}

\nocite{Wolff1989,Schlijper1989,Onsager1944,Wannier1950}

%


\newcommand\del[1]{\textcolor{magenta}{#1}}
\renewcommand\del[1]{}

\newcommand\add[1]{#1}

\renewcommand\thefigure{S\arabic{figure}}
\renewcommand\thetable{S\arabic{figure}}

\setcounter{section}{0}
\setcounter{figure}{0}
\setcounter{table}{0}


\title{Supplemental Material for \\
	Universal and accessible entropy estimation using a compression algorithm}

\date{\today}

\maketitle


\clearpage

\onecolumngrid

\chapter{\begin{center}\textbf{\uppercase {Supplemental Material for Universal and accessible entropy estimation using a compression algorithm}}\end{center}}

\twocolumngrid

\renewcommand\thefigure{S\arabic{figure}}
\renewcommand\thetable{S\arabic{figure}}

\setcounter{section}{0}
\setcounter{figure}{0}
\setcounter{table}{0}


\section{\label{appendixEntropyAndCompression}ON ENTROPY AND COMPRESSION ALGORITHMS}
Physical systems presented in this work are severely under-sampled compared to their available configurational space (\textit{e.g.}, $2^{4096}$ states for the Ising model on a square lattice with $64^2$ sites). In the text, we discuss a proof for the convergence of the size of a sequence encoded by the Lempel-Ziv algorithm \cite{Ziv1977}, to Shannon-entropy per symbol  \cite{Ziv1978}. It is worth noting that the analogy made there is between physical microstates and symbols of the processed sequence.  As a result, convergence is guaranteed only for an over-sampled sequence of the system's microstates. Using our method, the entropy estimate converges for a much smaller sample size [Figs. \ref{fig:one}(a-d)]. This is in accordance with conventional simulation methods (Metropolis Monte Carlo or molecular-dynamics) that produce reliable thermodynamical properties, even when under-sampled\cite{Frenkel2001,landau2014guide}.

We offer the following concise description of inner-workings of the current method, which may shed light on the reasons for which it works, and potential limits.

The probability $p_i$ of observing a system's i'th microstate, described as a set of variables $\{v_k\}$, can be broken down to the observation probability of sub-microstates (\textit{i.e.}, values of a sub-set of system variables), via the probability chain-rule: $p_i = P(v_1,...,v_D ) = P(v_{k+1},...,v_D  | v_1,...,v_k ) \cdot P(v_1,...,v_k )$, for an arbitrary index k. This decomposition can be continued to any arbitrary partitioning into sub-sets of variables. Note however that, in all systems, a correlation length ($l_c$) can be defined such that pairs of non-overlapping sub-microstates of size $l_c$, have a negligible correlation. Finally, microstates may be decomposed into uncorrelated sub-microstates of size $l\geq l_c$; in this case we end up removing the conditions in the chain-rule: $p_i = P(v_1,...,v_D ) = P(v_1,...,v_l ) \cdot ... \cdot P(v_{D-l+1},...,v_D )$, otherwise stated as:
$\log p_i = \log P (v_1,...,v_l)+...+\log P(v_{D-l+1},...,v_D)$.

Switching to the compression algorithm perspective, practical implementations find patterns of sequential bytes; these patterns can be viewed as sub-microstates. At each point during the compression process, patterns are matched to the history of data up to that point. A matched pattern is then replaced by a reference of size $\log d$ to data found a distance $d$ back. If one assumes an underlying Poisson process, and therefore an exponential distribution of distance to next observation, then the average distance is $\langle d \rangle = p^{-1}$; where $p$ is the probability of observing the current sub-microstate. The average size of compressed data amounts to $\langle \Sigma_p \log d_p \rangle$ per sampled configuration, where $p$ is the index for the current pattern within a conformation (\textit{i.e.}, sub-microstate), and $d_p$ is the distance to previous observation of the pattern. Empirically, due to the exponential distribution of $d_p$ one can replace $\langle \log d_p \rangle \approx \log \langle d_p \rangle$. Additionally, assuming for the size of the patterns $(l_p): l_p\geq l_c$, and using the statements above, we recover an average compressed size of $-\langle\log p_i \rangle$ per sampled configuration.

\section{\label{suppComparisonToPrevious}PREVIOUS STUDIES USING COMPRESSION ALGORITHMS}

A few previous studies used lossless compression on physical system simulations, taking different approaches than the one presented in this work \cite{Melchert2015,Vogel2012,Vogel2017,Martiniani2019}. For example, Melchert \& Hartmann \cite{Melchert2015} used compression to characterize the divergent correlation-time in the dynamics of an Ising model and commented on potential generalization for the detection of order-disorder phase transitions. Their approach is \del{very }useful for the characterization of divergent trends, as \del{occur, for example, with}in the case of correlation-time near a 2\textsuperscript{nd} order phase transition. However, as stated in their work, that approach does not lead to a universal estimate of \del{an ensemble's }entropy.

All \del{the }aforementioned studies \del{faced the challenge of converting}reduced multi-dimensional datasets to \del{strings}streams of \del{information}data that can be compressed by representing every time point with one value. Melchert \& Hartmann \cite{Melchert2015} compressed the time-series of an arbitrary spin in the 2d Ising lattice, and Vogel \textit{et al.} \cite{Vogel2017} compressed a time-series of a nematic order parameter. 

More recently, Martiniani \textit{et al.} used lossless compression to qualitatively evaluate the time-evolution of order in out-of-equilibrium systems \cite{Martiniani2019}. There, the exact nature of order is not easily defined and lossless compression \del{was found }was used to \del{identify}detect a dynamic phase transition. \del{In their work, the studied system is composed of a large lattice of sites, each with a finite number of values. }Their detection \del{of a dynamic phase transition }is achieved by monitoring the compressed-length of a full description of each configuration, as the system evolves in time. These studies did not attempt to quantify entropy; rather, they use divergent trends in information content, quantified by lossless compression, to detect a change in the system state.

Application of lossless compression to a single configuration for estimation of entropy, such as in the case of Martiniani \textit{et al.} \cite{Martiniani2019}, implicitly assumes an equivalence between a large system and an ensemble of smaller similar systems. Such equivalence is valid in systems like the Ising model or indeed any system with homogenous degrees of freedom and periodic boundaries. \del{Moreover, in these kinds of systems, we are interested in an intensive measure of entropy (entropy per site/volume, etc.), for a large system. Therefore, scaling-up the system results in a better entropy estimate. }However, \del{scaling-up}simulation of a large ensemble of complex \del{simulations}systems such as protein dynamics, although in-principle possible, is \add{impractial}\del{non-trivial and very computationally demanding}. Instead, we\del{one can} treat a sequence of independently sampled configurations as an \del{scaled-up system}ensemble and apply compression \del{to the entire ensemble}on its entirety. 

In this work, we are ultimately interested in systems which are heterogeneous\add{, without any obvious internal similarity (\textit{e.g.} symmetry)}\del{typically simulated as single units}. Examples of such systems include a stretched ideal chain and other linear chains such as proteins, DNA or RNA. If one was to apply a lossless compression algorithm to a single configuration of these systems, the resulting entropy value would typically be over-estimated since recurring patterns would be rare. Instead, in these systems, in order to achieve a valid estimate for entropy using lossless compression, we apply the compression to a set of many independent configurations of the system at once. This approach leads to pattern matches by the compression algorithm \textit{across the set of configurations}. In our work, we demonstrate that with the appropriate care, lossless compression algorithms can do more than detection of order and yield the \textit{entropy values} of a digitized physical realization.

\section{\label{AppendixDetailedCalculation}DETAILS OF CALCULATION OF $S_\mathrm{A}$}

\subsection{General scheme}
Practically, to quantify $S_\mathrm{A}$, each recorded simulation data was first encoded into a file as bytes. Encoding into bytes encompasses various choices of projection to lower dimensionality and coarse-graining, which will be discussed below. The encoded file is compressed using the Lempel-Ziv-Markov chain-Algorithm (LZMA), implemented in the open-source 7-Zip software (LZMA); the resulting compressed size in bytes is plugged into the calculation of $S_\mathrm{A}$, as described in the main text.

Concisely, to extract an entropy value from a digitized dataset one should perform the following steps:
\renewcommand{\labelenumi}{\alph{enumi}.}
\begin{enumerate}
	\item Choose appropriate degrees-of-freedom
	\item Reduce dimensionality (re-index coordinates)
	\item Coarse-grain
	\item Compress and rescale resulting file size to extract $S_\mathrm{A}$
	\item Evaluate preprocessing
\end{enumerate}

\subsubsection{Choose Appropriate Degrees-of-Freedom}
We consider a physical system with $D$ degrees of freedoms recorded at $N$ independently sampled configurations. The original recorded dataset is defined as the set of variables $\{x_i^t\}$, where $t=1...N$ and $i=1...D$. 
Often the native coordinate systems in which the system is recorded are not optimal for convergence of $S_\mathrm{A}$ under finite-sampling. For example, trivial degrees of freedom, such as whole-body translation and/or rotation typically do not affect configurational entropy (or free-energy). In such cases, a transformation to an orientation-aligned center-of-mass frame should be performed.  

\subsubsection{Reduce Dimensionality}
The LZMA algorithm, as well as any alternative, works by finding contiguous patterns of bytes which appear previously within the file. As a result, the algorithm only detects one-dimensional correlations, while local correlations within the data might be better represented in higher dimensionality. This problem has been conveniently analyzed before, where a reduction to one-dimension using a Hilbert space-filling curve (“Hilbert scan”) was found optimal to retain clustered correlations \cite{Moon2001}. In general, the choice of reduction from multi- to one- dimension affects convergence of $S_\mathrm{A}$. We have tested a multitude of schemes for this reduction and demonstrate several for the Ising model on a square lattice [Fig. \ref{fig:two}(a)].

\subsubsection{Coarse-Grain}
Compression algorithms are designed to minimally represent a dataset's alphabet (finite set of symbols, of which a sequence is composed). However, often data is recorded as continuous variables which contain insignificant digits that are effectively random and independent, due to noise or numerical inaccuracy. Such digits render the dataset's alphabet enormous. In principle, $S_\mathrm{A}$ asymptotically converges for any sized alphabet; however, for practical purposes, the required sample size increases dramatically. To treat the issue, we approximate a system's entropy by the entropy of a projected system with discretized degrees of freedom (\textit{i.e.}, coarse-graining), for which $S_\mathrm{A}$ converges at much smaller sample size.

Here again, many schemes for coarse-graining may be introduced which, after computation of $S_\mathrm{A}$, will produce an upper bound on the actual entropy. In this work, we coarse-grain the continuous degrees of freedom in the ideal-chain model, and in molecular dynamics trajectories, using the following scheme. For both systems, we have used a representation composed of angles. In the ideal chain these are bond angles relative to common axes, and in the protein, these are dihedral angles -- relative to surrounding amino acids. In either system, the maximal range of values taken up by the coordinates is conveniently known in advance to be $[-\pi,+\pi)$.

For each degree of freedom $x_i^t$, we generate a new integer variable $\tilde{x}_i^t=\lfloor n_\mathrm{s}⋅(x_i^t  + \Delta x_i )/R_i\rfloor$, which is effectively a rounding down to units of $R_i/n_\mathrm{s}$. Here, $\Delta x_i$ and $R_i$ are a shift and range, respectively, for the i'th degree of freedom, such that the mapped values would lay in the range $[0,n_\mathrm{s})$. In the current case, the shift is $+\pi$ and the range is preset to $2\pi$, but in general, these can be derived from the samples using, for example, the mean and some multiple of the standard deviation. The number of coarse-grained values $n_\mathrm{s}$ is optimized for minimal calculated $S_\mathrm{A}$ [Fig. \ref{fig:two}(d)]. Whether we started off with discrete degrees of freedom or continuous ones, the assessed discrete system invariably has maximal entropy range of $D \log n_\mathrm{s}$, where $D$ is the number of represented degrees of freedom.

\subsubsection{Compress and Rescale}
\add{For compression, LZMA version 16.04 was used (www.7-zip.org/download.html), with the following parameters: 3 literal context bits, 0 literal position bits, 2 position bits, 64 fast bytes, BT4 search tree, 4 hash bytes, and a dictionary size of 64 MB. These parameter values are the default compression settings of the 7Zip software, set to "maximal compression". The exact code used in this work can be provided upon request. All data streams compressed in this work were smaller than the set dictionary size. }

\begin{figure}
	\includegraphics[width=0.5\textwidth]{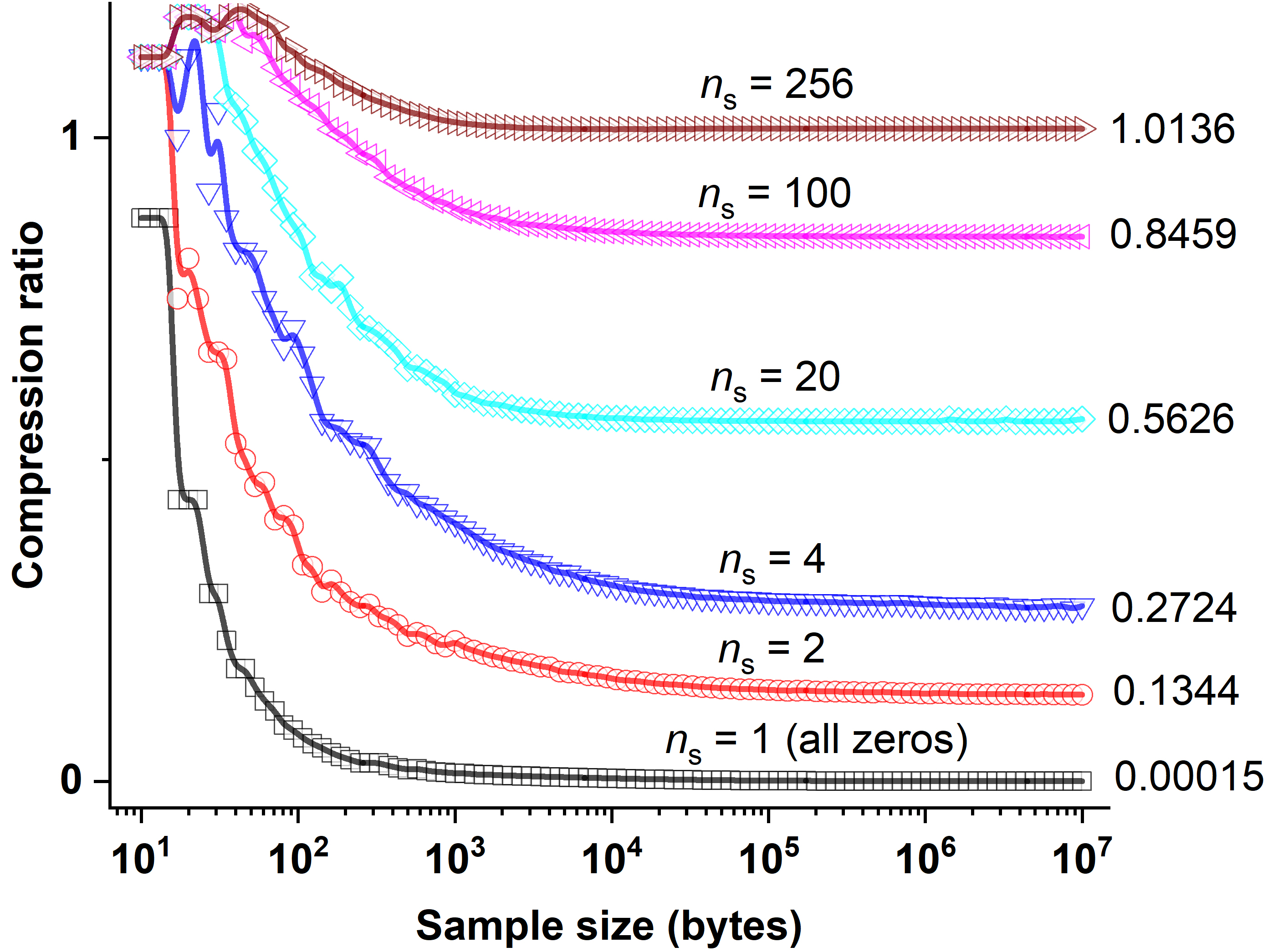}
	\caption{\label{fig:five} Convergence of compression ratio for varying alphabet size ($n_\mathrm{s}$). Data points represent the ratio between the compressed length and the original length for sequences of randomly generated integers from a uniform distribution. For compression, these numbers are stored each in a single byte, which can store an alphabet of at most 256 unique values. These compression ratios are used for the calibration of the entropy estimate in the maximal entropy state for a given choice of the number of coarse-grained represented values (the case of a single valued alphabet is always used as the minimal entropy state). Compression ratios for $10^7$ bytes are indicated to the right of each curve and due to practical limitations of the LZMA algorithm are slightly higher than the expected ratio $\log_2 n_\mathrm{s} / \log _2 256$, for each $n_\mathrm{s}$. For reference, the length of 5000 concatenated configurations of an ideal chain with 100 monomers is $~5\cdot 10^5$ bytes (1 byte per monomer per configuration).}
\end{figure}

As described in the main text, the final entropy estimate is derived by relation to the known entropies of the maximal and minimal entropy states (even if these are not reachable in the given system). These converge to a typical compression ratio for each choice of $n_\mathrm{s}$, as demonstrated in Fig. \ref{fig:five}. In principle,  representations  coarse-grained to $n_\mathrm{s}$ values require $\log_2 n_\mathrm{s}$ bits to be described, hence the expected asymptotic compression ratio would be $\log _2 n_\mathrm{s} / \log _2 256$ for representations occupying whole bytes (8 bits). However, due to practical limitations of the LZMA algorithm used here, the compression ratio converges to a value slightly larger than expected. This is normalized out in our definition of the incompressibility content $\eta = (C_\mathrm{d}-C_0)/(C_1-C_0 )$, as schematically shown in Figure \ref{fig:scheme}.

Practically, the maximal and minimal entropy states are derived by generating two additional datasets having the original dataset length. In the first, data over the entire phase-space is replaced with a single repeating symbol (\textit{e.g.}, zero). The result of compressing this "zeros" dataset  is defined as $C_0$ and is given in Fig. \ref{fig:five} by $n_\mathrm{s}=1$. In the second dataset, all the values are replaced with random symbols from the alphabet ($n_\mathrm{s}$) used to extract $C_\mathrm{d}$. The latter results with the compressed file size denoted by $C_1$. As described above, the entropy is calculated by $S_\mathrm{A}/k_\mathrm{B}=\eta D\log n_\mathrm{s}$.

\subsubsection{Evaluate Preprocessing}
Due to finite sampling, $S_\mathrm{A}$ is guaranteed to be an upper bound to the physical entropy of the system. Nonetheless, alternative choices in the preprocessing steps such as the level of coarse-graining and the choice of degrees-of-freedom may eliminate important correlations or introduce spurious effects challenging the compression algorithm's efficiency. It is therefore recommended to evaluate the entropy over various representations and to chose the one with the minimal valued $S_\mathrm{A}$.

\subsection{Implicit assumptions and spurious effects}
One should take note of the implicit physical assumptions made by a compression algorithm which processes bytes. Compression algorithms match patterns between any arbitrary pairs of locations, which implies translational symmetry (at least in the 1d representation). Many systems, like a polymer, protein, or indeed any finite system without a periodic boundary, do not have this symmetry. The implicit assumption of symmetries may lead to an under-estimation of entropy by $S_\mathrm{A}$ due to spurious matching of patterns. We do not currently observe such an effect and can only assume that matches between independent regions in our tested systems are much less likely than matches between the same region in different instances.
\subsection{Calculation of $S_\mathrm{A}$ for the finite states system}
States sampled from the finite state system were laid out consecutively in a file, with each byte holding the index of the currently selected state.  For the entropy estimate, $D=1$ and $n_\mathrm{s}$ is assigned the number of energy levels.We construct and compress the zero dataset ($C_0$) as a file of identical size to the recorded files above, with all samples set to the value 0. A random dataset ($C_1$) is created similarly, with every sample chosen randomly and uniformly from the available alphabet.
\subsection{Calculation of $S_\mathrm{A}$ for the Ising model}
The Ising model consists of a $\pm 1$ value per site, which we represent by a single binary digit (a bit). Spin sites on the 2d square lattice are laid out in a 1d sequence either row-by-row, in a spiral (first row, then last column without overlapping site, spiraling inwards in clockwise order), or ordered by a Hilbert space-filling curve \cite{Moon2001}. Site values (bits) are laid out in a file either consecutively, filling every byte with the values of 8 sites (8 bits), or 1 site value per byte. Also, we used an intermediate filling of bytes with 3 sites per byte, where the two additional values were taken from the sites above and to the right, regardless of the order in which the sites are laid out.
For the triangular lattice, we use a Hilbert scan with 3 sites per byte, as described above. In practice, we implement the triangular lattice as a square lattice with two additional diagonal bonds, so scans regard the site positions as for the square lattice.
For the final estimate, $D$ is assigned the number of sites in the lattice and $n_\mathrm{s}$ = 2.
The zero dataset ($C_0$) is generated with configurations of equal size and number to the sampled systems above, with all spin values set to -1. The random dataset ($C_1$) is produced by setting spins uniformly and randomly to $\pm 1$ and representing as described above.
\subsection{Calculation of $S_\mathrm{A}$ for the ideal chain}
\begin{figure}
	\includegraphics[width=0.36\textwidth]{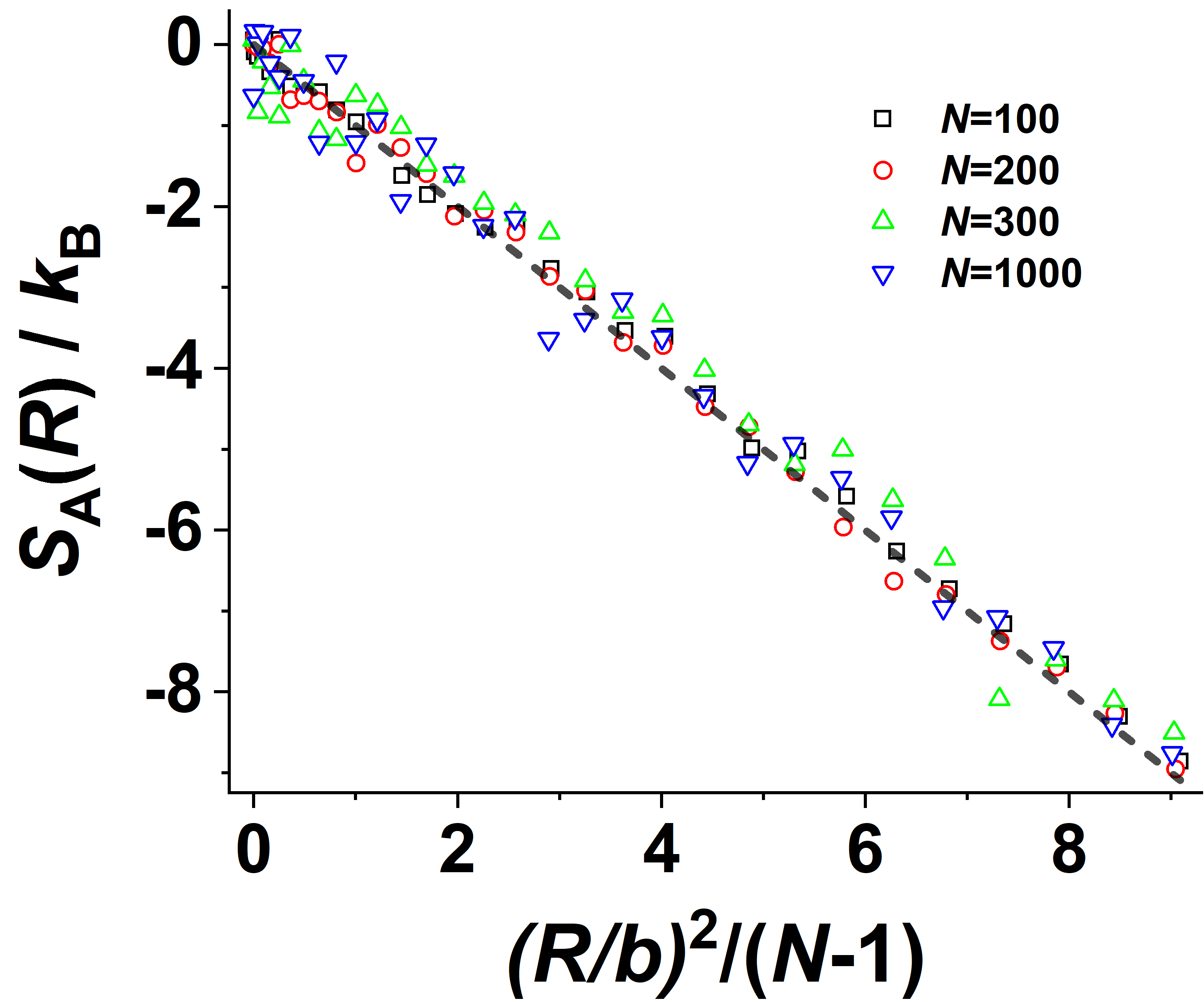}
	\caption{\label{fig:six}Entropy estimates for the ideal chain, normalized by number of bonds. Data points from Fig. \ref{fig:one}(e) of entropy estimates for an ideal chain of varying length ($N$) and constant bond length (b), with increasing fixed end-to-end distance ($R$), normalized by number of bonds ($N-1$). Data points collapse on the theoretically expected entropy $S=-R^2/b^2 (N-1)$ depicted by the dashed line. }
\end{figure}

Our simulations of ideal-chains consisting of $D=2\cdot N$ coordinates were conducted and recorded with 64-bit precision floating point numbers. We tested several representations for the chain to be compressed, including a naive representation with cartesian coordinates. For the ideal chain, a representation as a list of angles for the steps between monomers (\textit{i.e.}, bond angles) resulted in the lowest entropy estimates and was therefore used for analysis. This representation effectively decorrelates the monomers and therefore produces the lowest entropy estimates, which is also robust to large deviations from the optimal $n_\mathrm{s}$, as observed by the shallow minima [Fig. \ref{fig:two}(d)]. Minimal $S_\mathrm{A}$ is consistently around $n_\mathrm{s}=170$ coarse-grained angles. The results presented here [Figs. \ref{fig:one}(e), \ref{fig:two}(d) and \ref{fig:six}] were derived using this representation \add{and also demonstrate that the compression algorithm effectively captures entropic contributions of individual variables, even when uncorrelated}. For calibration, we generate the zero ($C_0$) and random ($C_1$) files as before, with configurations equal in size and number to recorded data.
In Fig. \ref{fig:six} we demonstrate the collapse of all our data points to a single curve described by $S=-R^2/b^2 (N-1)$ when the x axis is normalized by $1/(N-1)$.

\begin{figure}
	\includegraphics[width=0.42\textwidth]{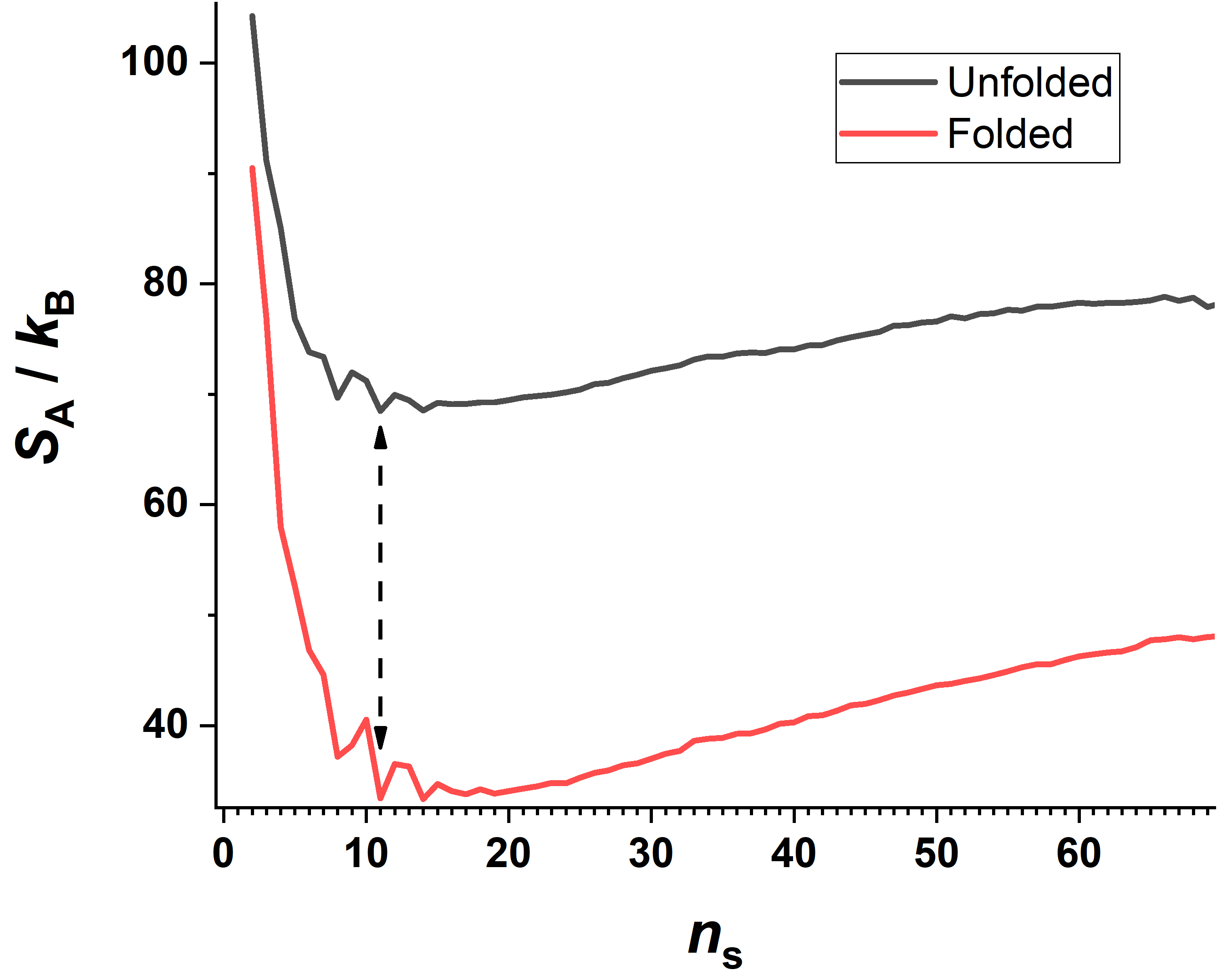}
	\caption{\label{fig:seven} Evaluation of entropy for varying number of coarse-grained angles ($n_\mathrm{s}$) for Villin headpiece. $S_\mathrm{A} (n_\mathrm{s})$ is calculated for 4,000 frames from either unfolded or folded configurations as assigned in this work, resampled every 100'th frame (see main text). Lowest $S_\mathrm{A}$ is calculated in both ensembles with $n_\mathrm{s}=11$, indicated by the arrows. We notice that the difference in entropies between the folded and unfolded states remains almost constant over a large range of $n_\mathrm{s}$.}
\end{figure}
\begin{figure}
	\includegraphics[width=0.35\textwidth]{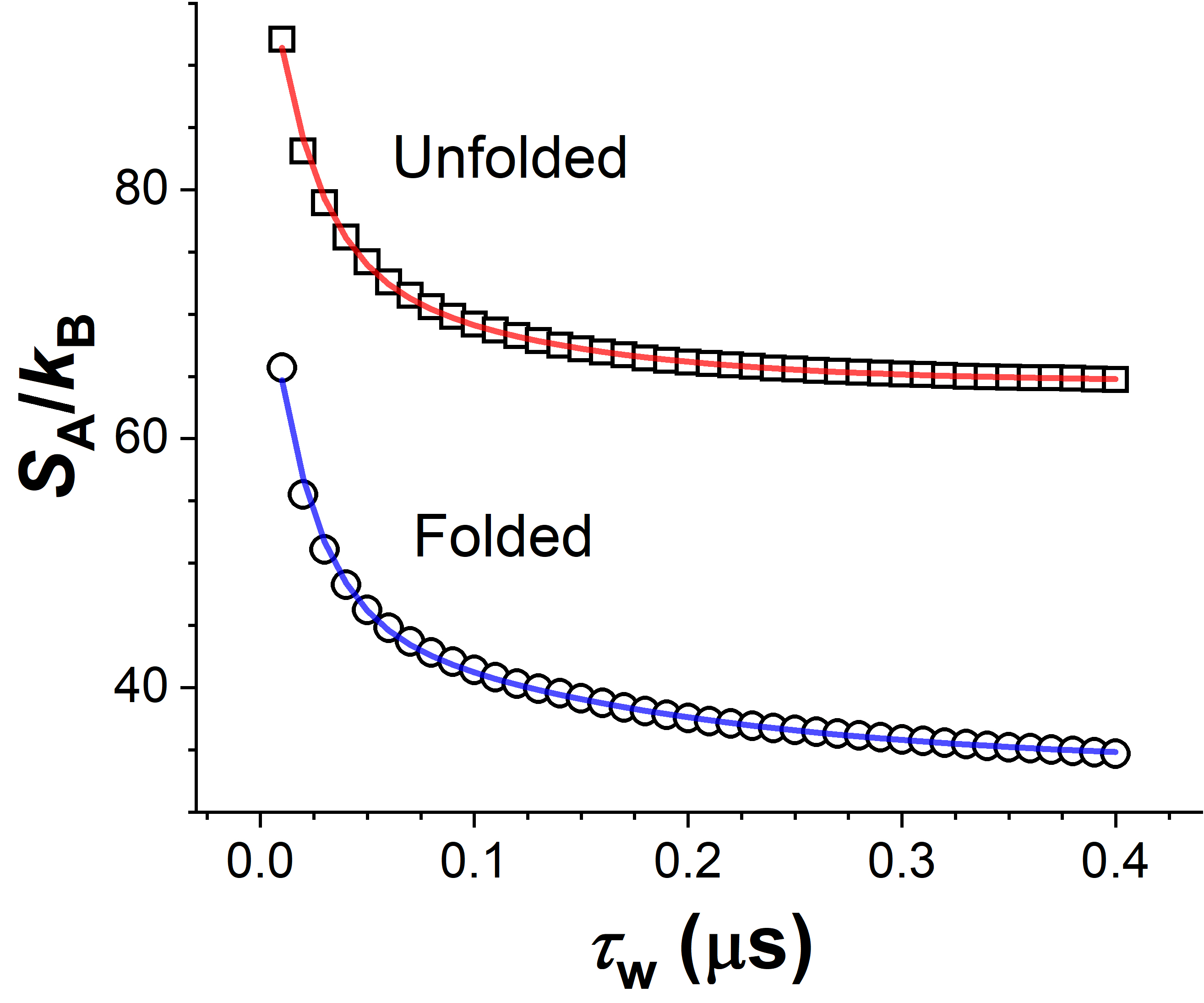}
	\caption{\label{fig:eight} Convergence of $S_\mathrm{A}$ with sampled window size $\tau _w$. $S_\mathrm{A} (\tau _w)$ is calculated for varying windows size, and averaged over 100 windows preassigned by the transition-based-assignment \cite{Piana2012} as folded (circles) and unfolded (squares). The calculated $S_\mathrm{A} (\tau _w)$ points are fit with a double exponential function$a+b_1 \cdot \exp (-\tau_\mathrm{w} / \tau_1) + b_2 \cdot \exp (-\tau_\mathrm{w} / \tau_2)$, which results with $\tau_1 \approx 0.02 \mathrm{\mu s}$ for both and $\tau_2=0.160 \mathrm{\mu s}$ and $0.100 \mathrm{\mu s}$ for the folded and unfolded frames respectively (solid lines).}
\end{figure}

\begin{figure*}
	\includegraphics[width=0.8\textwidth]{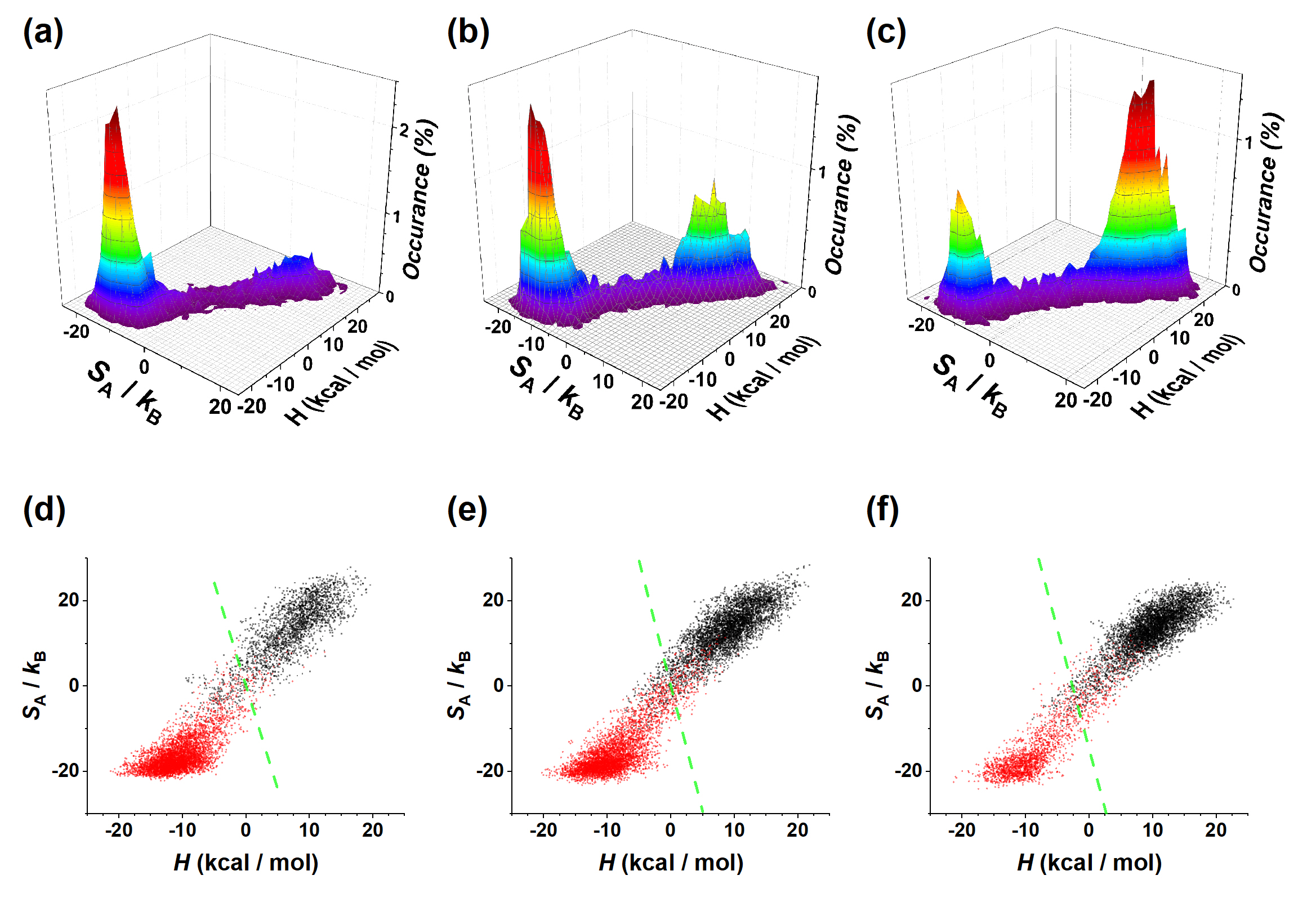}
	\caption{\label{fig:nine} Compression based Entropy ($S_\mathrm{A}$) -- Enthalpy ($H$) diagrams. (a-c) Enthalpy -- entropy population diagrams calculated from enthalpy assignment during the simulations and entropy assigned from compression method. Two well-defined states (folded and unfolded) of high and low entropy are clearly demonstrated. (d-f) Scatter plots of the same analysis colored with preassigned folded (red) and unfolded (black), according to the transition-based-assignment \cite{Piana2012}. A dashed green line divides the reassigned folded and unfolded states, as explained in the methods. (a,d) Simulation at T=360K. (b,e) Simulation at T=370K. (c,f) Simulation at T=380K.}
\end{figure*}
\subsection{Calculation of $S_\mathrm{A}$ for the Villin headpiece protein fragment}
We treat the coordinates for each $\mathrm{C_\alpha}$ carbon essentially as for the Ideal chain. Here, however, we have two dihedral angles per site ($\phi, \psi$, $D=66$). We choose to lay out a sequence of concatenated pairs of angles ($\phi_1, \psi_1, \phi_2, \psi_2, ... $) rather than concatenating first one angle and then the other ($\phi_1, \phi_2, ..., \psi_1, \psi_2, ... $) as this results in slightly lower $S_\mathrm{A}$ values. 
For $\Delta S_\mathrm{A}$ values (see Table \ref{tab:table1} in the main text), frames assigned to either folded or unfolded state are collected for processing. We tested for correlation time between frames, using the convergence of compression ratio [see for example Fig. \ref{fig:two}(c)] and found correlation times in the range of 20-30 frames for collected frames of either folded or unfolded assignments. Therefore, every 100'th frame was kept for further processing. We observe lowest $S_\mathrm{A}$ values for $n_\mathrm{s}=11$ coarse-grained values per coordinate, in either the folded or unfolded configurational ensembles (Fig. \ref{fig:seven}).

\begin{table}[b] \caption{\label{tab:table1}
		Thermodynamics estimation derived from MD simulations of Villin headpiece at three  temperatures ($T$). Folding free energy estimation was calculated from the ratio of folded and unfolded states either pre-assigned by Piana \textit{et al.} \cite{Piana2012}  ($\Delta G_\mathrm{f}$) or assigned by our method  ($\Delta G_\mathrm{A}$). Folding enthalpy ($\Delta H_\mathrm{f}$) was calculated from the difference in average force-field energy in the pre-assigned data \cite{Piana2012}. Entropy difference between folded to unfolded ensembles is compared between the transition-based assignment method ($T\Delta S_\mathrm{f}=\Delta H_\mathrm{f}-\Delta G_\mathrm{f}$) and the compression-based entropy estimations ($\Delta S_\mathrm{A}$), as detailed in the text. All energy terms are given in $\mathrm{\frac{kcal}{mol}}$ units.}
	\begin{ruledtabular}
		\begin{tabular}{ l | c  c  c  c c }
			\textrm{$T (K)$}&
			\textrm{$\Delta H_\mathrm{f}$} &
			\textrm{$\Delta G_\mathrm{f}$}&
			\textrm{$T\Delta S_\mathrm{f}$}&
			\textrm{$\Delta G_\mathrm{A}$}&
			\textrm{$T\Delta S_\mathrm{A}$}\\
			\colrule
			360 & -17.6 & -0.6 & 17.0 & -0.64	& 25.9\\
			370	&	-18.0& 0.0  & 18.0 & -0.02 &	25.4\\
			380	& -21.2 &  0.7	 &  21.9 &  0.81	& 25.9\\	
		\end{tabular}	
	\end{ruledtabular}
\end{table}

We generate the zero ($C_0$) and random ($C_1$) files as before, with configurations equal in size and number to processed recorded data. We generate the random file from uniformly random $n_\mathrm{s}$ dihedral angles per coordinate.
For the sliding-window $S_\mathrm{A}$ values (Figs. \ref{fig:protein} and \ref{fig:eight}), we obtain a reasonable compromise between convergence of $S_\mathrm{A}$ and time-resolution with a window of 2,000 consecutive frames ($\tau_\mathrm{w}=0.4 \mathrm{\mu s}$, see Fig. \ref{fig:eight}), independent of choice of $n_\mathrm{s}$. Our  $S_\mathrm{A}$ estimate is evaluated for discretized windows starting every 200 frames (90\% overlap). Using $n_\mathrm{s}=24$ results in maximal discrepancy between $S_\mathrm{A}$ values for either folded or unfolded states and was therefore used for the $S_\mathrm{A}-H$ diagrams [Figs. \ref{fig:protein}(b) and \ref{fig:eight}] instead of $n_\mathrm{s}=11$ which was used for the ensemble entropy estimates in Table \ref{tab:table1}. For these enthalpy-entropy diagrams, we averaged per-frame enthalpy values provided with the recorded simulation, over the same windows defined above.

For each simulated temperature, a line crossing the valley between clustered events was optimized as follows. We interpolate values of a two-dimensional $S_\mathrm{A}-H$ histogram along an optimized line and minimize the line integral over sampled values. Final optimized lines are given by $S_\mathrm{A}=aH+b$ with the parameters: ($a=-4.84, b=0$), ($a=-5.88, b=-0.01$), ($a=-5.61, b=-14.9$) for temperatures 360K, 370K, 380K respectively. These optimized classification line borders are shown in Fig. \ref{fig:nine}. After assigning windows above (below) the line as unfolded (folded), we compared to per-frame assignments given to us with the recorded simulation achieved with the “transition-based-assignment” \cite{Piana2012}, which are averaged over windows. The agreement reached 96.3\%, 95.6\% and 95.3\% for temperatures 360K, 370K, 380K respectively.

\section{\label{appendixSimulation}SIMULATIONS}

\subsection{\label{app:subsecA}Finite states simulations}
A discrete state system was defined consisting of the following four energy-levels: $\epsilon,2\epsilon,70\epsilon,80\epsilon$, with $\epsilon$ being an arbitrary energy scale. States were sampled from their Boltzmann distribution $p_i = \frac{ \exp^{-\varepsilon_i/\epsilon T}}{ \Sigma_j \exp^{-\varepsilon_j/\epsilon T} }$ with $p_i$ the probability for the i'th energy-level, $\varepsilon_i$ its energy, and $T$ the dimensionless simulation temperature. For each temperature, $7\cdot 10^6$ states were sampled and recorded.

\subsection{\label{app:subsecB} Ising model simulations}

Ising spin-half simulations were conducted using the Metropolis Monte-Carlo (MC) algorithm at various temperatures. An interaction potential was defined between each nearest neighboring pair of sites $(i,j): -1/2 \cdot J\sigma_i \sigma_j$, with $\sigma_i$ a $\pm 1$ valued spin site and periodic boundary conditions. Sites were put either on a square or a triangular lattice, with the number of sites $L^2=64^2$ for the ferromagnetic ($J=+1$) and $16^2$ for the antiferromagnetic ($J=-1$) models. The MC step involves either a “single site flip” or a “cluster flip”, depending on the temperature and coupling parameter $J$. A single site flip involves a randomly picked site which is flipped with probability $\exp^{-\Delta E/k_\mathrm{B} T}$ for $\Delta E>0$ energy difference caused by the flip, and with probability 1 for $\Delta E\leq 0$; $T$ being the simulation temperature. The cluster flip involves randomly picking a site, and repeatedly growing a cluster onto neighboring sites. Cluster growth is considered through each outward bond from sites on the cluster's interface, onto sites with an identical sign, and with probability $1-\exp^{-2J/k_\mathrm{B} T}$, as described by U. Wolff \cite{Wolff1989}. For the antiferromagnetic triangular lattice, only single site flips were used.

\begin{figure}
	\includegraphics[width=0.35\textwidth]{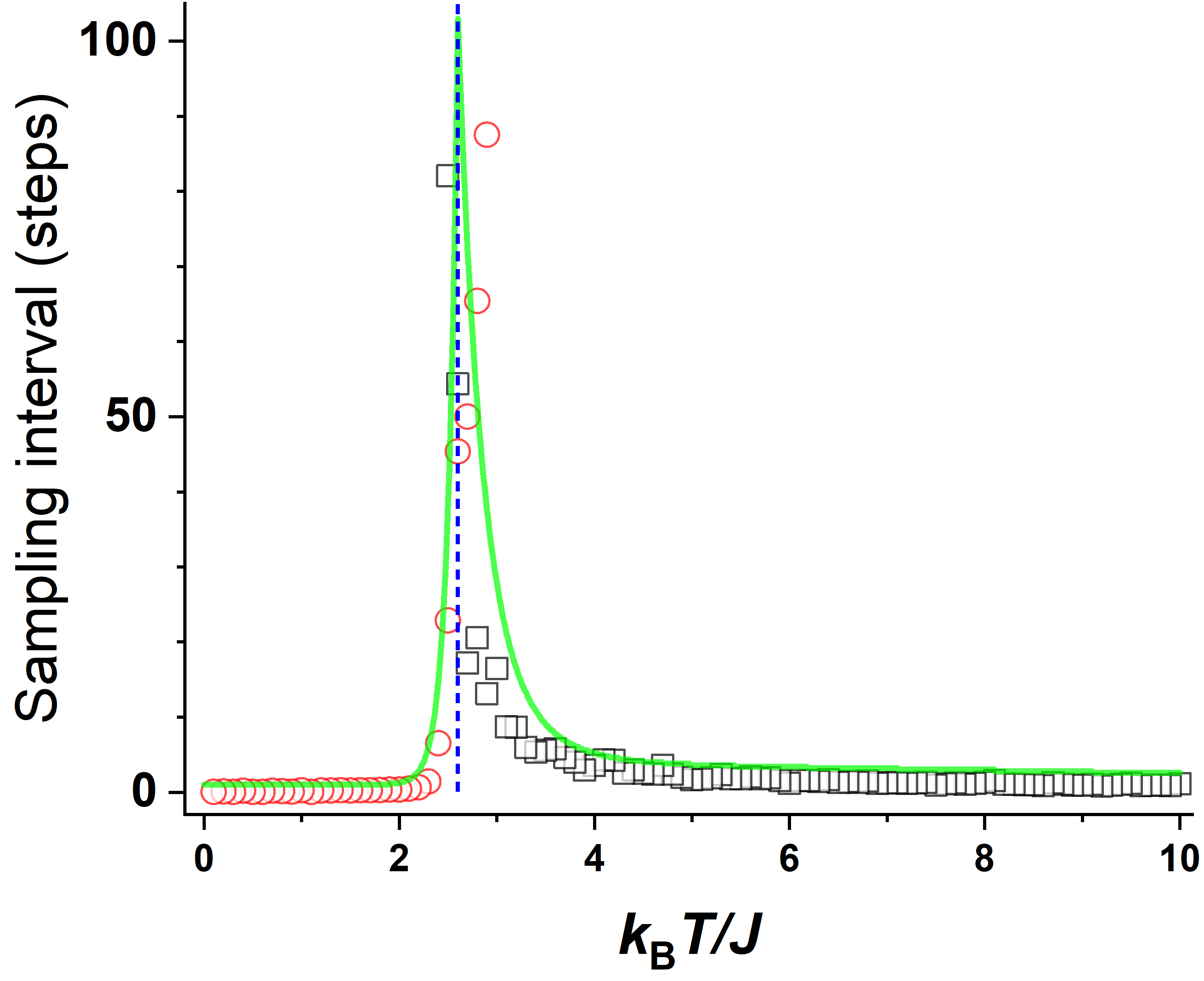}
	\caption{\label{fig:four} Correlation times and sampling plan for the Ising model on a 64 by 64 square lattice. Correlations times were calculated for the mean spin value fluctuations, with either single-flip steps ¬(squares) or cluster-flip steps (circles). Sampling intervals are stated in $L^2$ steps for single-flips and single steps for cluster-flips. The temperature of transition to the cluster-flip steps ($T=2.6  k_\mathrm{B}/J$) is plotted (dashed line). For each temperature, the simulation is sampled at intervals indicated by the sampling plan (green line, scaled by $1/5$).}
\end{figure}

Correlation times were calculated by fitting the autocorrelated fluctuations of mean spin value, with an exponential function; this was done for each simulated temperature. The final sampling plan defined sampling intervals exceeding the correlation times 5-fold, at each temperature (Fig. \ref{fig:four}). Based on correlation times for either single or cluster flips, cluster flips were used with probability 0.5, below $T=2.6 k_\mathrm{B}/J$ for the ferromagnetic model on a square lattice and $T=4.1 k_\mathrm{B}/J$ on a triangular lattice. Simulations were started in a random configuration and run consecutively at decreasing temperature, with a recording of full system microstates every sampling interval, until 5,000 sampled configurations. Simulations were verified by comparing entropy quantified using the standard cluster variation method \cite{Schlijper1989}, to the analytical entropy derived by Onsager \cite{Onsager1944} for the square and Wannier \cite{Wannier1950} for the triangular lattices.

\subsection{\label{app:subsecC}Two-dimensional ideal chain simulations}
Ideal chain simulations were conducted using MC at varying end-to-end distance $R$, in a lattice-free setup. The number of monomers ($N _\mathrm{m}=D/2$) along the chain was set to either 100, 200, 300 or 1000, and the bond length ($b$) was constantly set to 1. At each MC step, two sites along the chain are randomly picked, and the chain in between is flipped about the line connecting the two sites. Minimal and maximal distance between the picked sites is set to 3 and $N _\mathrm{m}/2$ respectively. The step occurs at probability 1, since no potential is defined, and the step preserves bond lengths. The simulation algorithm is verified by running with an additional step of randomly reorienting the edge bonds with probability 0.1, to get a free ideal chain. The observed end-to-end vector populations fit the expected Gaussian distribution.

\begin{table*}[t] \caption{\label{tab:tableAccuracy}
		Errors and uncertainties calculated from 11 independent entropy evaluations
	}
	\begin{ruledtabular}
		\begin{tabular}{ l | c  c  c  c c }
			Model & \# configurations  & Entropy scale & $\left<\delta S_\mathrm{A}(T)\right>_T$ & $\mathrm{max} \left\{ \delta S(T) \right\} _T $ & $\mathrm{max}\{\mathrm{SD}(S_\mathrm{A})\}_T $ \\
			& &  ($k_\mathrm{B}$) &  ($k_\mathrm{B}$) &  ($k_\mathrm{B}$) &  ($k_\mathrm{B}$) \\
			\colrule
			4 state model & $10^7$ & $\mathrm{log}(4)=1.6$ & 0.013 & 0.06 & 0.009 \\
			Ising on a square lattice (64 x 64) & $10^4$ & $\mathrm{log}(2)=0.69$ & 0.017 & 0.04 & 0.0003 \\
			Ideal chain $N=100$ & 5000 & $R^2 / (b^2 N)=9$ & 0.03 & 0.09 & 0.08 \\
			Ideal chain $N=1000$ & 5000 & $R^2 / (b^2 N)=9$ & 0.07 & 0.14 & 0.14 \\
		\end{tabular}	
	\end{ruledtabular}
\end{table*}

We calculate correlation times by fitting an exponential function to the autocorrelated fluctuations of the radius-of-gyration $R _\mathrm{g}^2 = \Sigma_i (r_i-\langle r \rangle_i)^2$, calculated over the chain coordinates $r_i$; this is done for each simulated $R$. The final sampling plan defines sampling intervals exceeding correlation times 5-fold, at each $R$. Simulations are started from a “zig-zag” configuration, equilibrated for 10,000 steps, and we record the chain's coordinates every sampling interval and store 5,000 sampled configurations per choice of $R$.

\subsection{\label{app:subsecD}Villin headpiece protein fragment}
Recorded simulation data for the Villin headpiece protein fragment \cite{Piana2012} was made available to us by the group of Dr. Shaw. The recorded simulation data processed in this work consists of a full-atom description of the 35 amino acids of the protein. The full-atom description contains hundreds of atoms, which are then reduced to a description of dihedral angles (such as in Ramachandran plots). Since dihedral angles refer to orientation changes between amino acids, they are undefined at the first and last ones. Each protein configuration is reduced to 2 dihedral angles per each of the 33 non-edge amino acids -- 66 coordinates in total. Data used for analysis is from the “Nle/Nle” mutant, simulated at 360K, 370K, and 380K.

\section{\label{appendixAccuracy}$S_\mathrm{A}$ ACCURACY}
We estimated the error range and uncertainties in our entropy calculation by repeating the compression-based assessment over 11 newly generated and uncorrelated datasets for several different systems. 
We define the uncertainty  $\delta S_\mathrm{A}(T)=\overline{S_\mathrm{A}}(T)-S(T)$. Here, $S(T)$ is the analytical entropy value for a given temperature ($T$), and $\overline{S_\mathrm{A}}(T)$ is the mean of entropy estimation over the realizations. In addition, error range is calculated by the standard deviation (SD) out of the realizations. 
In Table \ref{tab:tableAccuracy}, we show the average and maximum uncertainties, and the maximum SDs over the entire temperature range under study. For the ideal chain case, T is replaced with end-to-end distance (R) and the analytical model $S(R)=S_0-R^2/(b^2 (N-1))$ was fitted for its constant $S_0$. 

When trying to evaluate how accurate our estimation is, in the most general way, we notice the problem that relative error ($\delta S_\mathrm{A}/S$) can be misleading when $S \rightarrow 0$. There, even a very good estimation of the entropy will result in divergence of the relative error.  Moreover, for continuous variables as in cases of the ideal chain and protein folding, only entropy difference from a reference should be used. However, entropy error ranges and uncertainties are orders of magnitude smaller than the entropy scale of the problem, giving us confidence using the compression-based scheme to evaluate entropy.


\begin{thebibliography}{36}%
	\makeatletter
	\providecommand \@ifxundefined [1]{%
		\@ifx{#1\undefined}
	}%
	\providecommand \@ifnum [1]{%
		\ifnum #1\expandafter \@firstoftwo
		\else \expandafter \@secondoftwo
		\fi
	}%
	\providecommand \@ifx [1]{%
		\ifx #1\expandafter \@firstoftwo
		\else \expandafter \@secondoftwo
		\fi
	}%
	\providecommand \natexlab [1]{#1}%
	\providecommand \enquote  [1]{``#1''}%
	\providecommand \bibnamefont  [1]{#1}%
	\providecommand \bibfnamefont [1]{#1}%
	\providecommand \citenamefont [1]{#1}%
	\providecommand \href@noop [0]{\@secondoftwo}%
	\providecommand \href [0]{\begingroup \@sanitize@url \@href}%
	\providecommand \@href[1]{\@@startlink{#1}\@@href}%
	\providecommand \@@href[1]{\endgroup#1\@@endlink}%
	\providecommand \@sanitize@url [0]{\catcode `\\12\catcode `\$12\catcode
		`\&12\catcode `\#12\catcode `\^12\catcode `\_12\catcode `\%12\relax}%
	\providecommand \@@startlink[1]{}%
	\providecommand \@@endlink[0]{}%
	\providecommand \url  [0]{\begingroup\@sanitize@url \@url }%
	\providecommand \@url [1]{\endgroup\@href {#1}{\urlprefix }}%
	\providecommand \urlprefix  [0]{URL }%
	\providecommand \Eprint [0]{\href }%
	\providecommand \doibase [0]{http://dx.doi.org/}%
	\providecommand \selectlanguage [0]{\@gobble}%
	\providecommand \bibinfo  [0]{\@secondoftwo}%
	\providecommand \bibfield  [0]{\@secondoftwo}%
	\providecommand \translation [1]{[#1]}%
	\providecommand \BibitemOpen [0]{}%
	\providecommand \bibitemStop [0]{}%
	\providecommand \bibitemNoStop [0]{.\EOS\space}%
	\providecommand \EOS [0]{\spacefactor3000\relax}%
	\providecommand \BibitemShut  [1]{\csname bibitem#1\endcsname}%
	\let\auto@bib@innerbib\@empty
	\bibitem [{\citenamefont {Dror}\ \emph {et~al.}(2012)\citenamefont {Dror},
		\citenamefont {Dirks}, \citenamefont {Grossman}, \citenamefont {Xu},\ and\
		\citenamefont {Shaw}}]{Dror2012}%
	\BibitemOpen
	\bibfield  {author} {\bibinfo {author} {\bibfnamefont {R.~O.}\ \bibnamefont
			{Dror}}, \bibinfo {author} {\bibfnamefont {R.~M.}\ \bibnamefont {Dirks}},
		\bibinfo {author} {\bibfnamefont {J.}~\bibnamefont {Grossman}}, \bibinfo
		{author} {\bibfnamefont {H.}~\bibnamefont {Xu}}, \ and\ \bibinfo {author}
		{\bibfnamefont {D.~E.}\ \bibnamefont {Shaw}},\ }\href {\doibase
		10.1146/annurev-biophys-042910-155245} {\bibfield  {journal} {\bibinfo
			{journal} {Annual Review of Biophysics}\ }\textbf {\bibinfo {volume} {41}},\
		\bibinfo {pages} {429} (\bibinfo {year} {2012})}\BibitemShut {NoStop}%
	\bibitem [{\citenamefont {Piana}\ \emph {et~al.}(2012)\citenamefont {Piana},
		\citenamefont {Lindorff-Larsen},\ and\ \citenamefont {Shaw}}]{Piana2012}%
	\BibitemOpen
	\bibfield  {author} {\bibinfo {author} {\bibfnamefont {S.}~\bibnamefont
			{Piana}}, \bibinfo {author} {\bibfnamefont {K.}~\bibnamefont
			{Lindorff-Larsen}}, \ and\ \bibinfo {author} {\bibfnamefont {D.~E.}\
			\bibnamefont {Shaw}},\ }\href {\doibase 10.1073/pnas.1201811109} {\bibfield
		{journal} {\bibinfo  {journal} {Proceedings of the National Academy of
				Sciences}\ }\textbf {\bibinfo {volume} {109}},\ \bibinfo {pages} {17845}
		(\bibinfo {year} {2012})}\BibitemShut {NoStop}%
	\bibitem [{\citenamefont {Piana}\ \emph {et~al.}(2014)\citenamefont {Piana},
		\citenamefont {Klepeis},\ and\ \citenamefont {Shaw}}]{Piana2014}%
	\BibitemOpen
	\bibfield  {author} {\bibinfo {author} {\bibfnamefont {S.}~\bibnamefont
			{Piana}}, \bibinfo {author} {\bibfnamefont {J.~L.}\ \bibnamefont {Klepeis}},
		\ and\ \bibinfo {author} {\bibfnamefont {D.~E.}\ \bibnamefont {Shaw}},\
	}\href {\doibase https://doi.org/10.1016/j.sbi.2013.12.006} {\bibfield
		{journal} {\bibinfo  {journal} {Current Opinion in Structural Biology}\
		}\textbf {\bibinfo {volume} {24}},\ \bibinfo {pages} {98 } (\bibinfo {year}
		{2014})}\BibitemShut {NoStop}%
	\bibitem [{\citenamefont {Kofke}(2005)}]{Kofke2005}%
	\BibitemOpen
	\bibfield  {author} {\bibinfo {author} {\bibfnamefont {D.~A.}\ \bibnamefont
			{Kofke}},\ }\href {\doibase 10.1016/j.fluid.2004.09.017} {\bibfield
		{journal} {\bibinfo  {journal} {Fluid Phase Equilibria}\ }\textbf {\bibinfo
			{volume} {228-229}},\ \bibinfo {pages} {41} (\bibinfo {year}
		{2005})}\BibitemShut {NoStop}%
	\bibitem [{\citenamefont {Landau}\ and\ \citenamefont
		{Binder}(2014)}]{landau2014guide}%
	\BibitemOpen
	\bibfield  {author} {\bibinfo {author} {\bibfnamefont {D.~P.}\ \bibnamefont
			{Landau}}\ and\ \bibinfo {author} {\bibfnamefont {K.}~\bibnamefont
			{Binder}},\ }\href@noop {} {\emph {\bibinfo {title} {A guide to Monte Carlo
				simulations in statistical physics}}},\ \bibinfo {edition} {4th}\ ed.\
	(\bibinfo  {publisher} {Cambridge university press},\ \bibinfo {year}
	{2014})\BibitemShut {NoStop}%
	\bibitem [{\citenamefont {Frenkel}\ and\ \citenamefont
		{Smit}(2001)}]{Frenkel2001}%
	\BibitemOpen
	\bibfield  {author} {\bibinfo {author} {\bibfnamefont {D.}~\bibnamefont
			{Frenkel}}\ and\ \bibinfo {author} {\bibfnamefont {B.}~\bibnamefont {Smit}},\
	}\href@noop {} {\emph {\bibinfo {title} {{Understanding molecular simulation:
					from algorithms to applications}}}},\ Vol.~\bibinfo {volume} {1}\ (\bibinfo
	{publisher} {Academic press},\ \bibinfo {year} {2001})\BibitemShut {NoStop}%
	\bibitem [{\citenamefont {Kubelka}\ \emph {et~al.}(2006)\citenamefont
		{Kubelka}, \citenamefont {Chiu}, \citenamefont {Davies}, \citenamefont
		{Eaton},\ and\ \citenamefont {Hofrichter}}]{Kubelka2006}%
	\BibitemOpen
	\bibfield  {author} {\bibinfo {author} {\bibfnamefont {J.}~\bibnamefont
			{Kubelka}}, \bibinfo {author} {\bibfnamefont {T.~K.}\ \bibnamefont {Chiu}},
		\bibinfo {author} {\bibfnamefont {D.~R.}\ \bibnamefont {Davies}}, \bibinfo
		{author} {\bibfnamefont {W.~A.}\ \bibnamefont {Eaton}}, \ and\ \bibinfo
		{author} {\bibfnamefont {J.}~\bibnamefont {Hofrichter}},\ }\href {\doibase
		10.1016/j.jmb.2006.03.034} {\bibfield  {journal} {\bibinfo  {journal}
			{Journal of Molecular Biology}\ }\textbf {\bibinfo {volume} {359}},\ \bibinfo
		{pages} {546} (\bibinfo {year} {2006})}\BibitemShut {NoStop}%
	\bibitem [{\citenamefont {Hansen}\ and\ \citenamefont {van
			Gunsteren}(2014)}]{Hansen2014}%
	\BibitemOpen
	\bibfield  {author} {\bibinfo {author} {\bibfnamefont {N.}~\bibnamefont
			{Hansen}}\ and\ \bibinfo {author} {\bibfnamefont {W.~F.}\ \bibnamefont {van
				Gunsteren}},\ }\href {\doibase 10.1021/ct500161f} {\bibfield  {journal}
		{\bibinfo  {journal} {Journal of Chemical Theory and Computation}\ }\textbf
		{\bibinfo {volume} {10}},\ \bibinfo {pages} {2632} (\bibinfo {year}
		{2014})}\BibitemShut {NoStop}%
	\bibitem [{\citenamefont {Buchete}\ and\ \citenamefont
		{Hummer}(2008)}]{Buchete2008}%
	\BibitemOpen
	\bibfield  {author} {\bibinfo {author} {\bibfnamefont {N.-V.}\ \bibnamefont
			{Buchete}}\ and\ \bibinfo {author} {\bibfnamefont {G.}~\bibnamefont
			{Hummer}},\ }\href {\doibase 10.1021/jp0761665} {\bibfield  {journal}
		{\bibinfo  {journal} {The Journal of Physical Chemistry B}\ }\textbf
		{\bibinfo {volume} {112}},\ \bibinfo {pages} {6057} (\bibinfo {year}
		{2008})}\BibitemShut {NoStop}%
	\bibitem [{\citenamefont {Shannon}(1948)}]{Shannon1948}%
	\BibitemOpen
	\bibfield  {author} {\bibinfo {author} {\bibfnamefont {C.~E.}\ \bibnamefont
			{Shannon}},\ }\href {\doibase 10.1002/j.1538-7305.1948.tb00917.x} {\bibfield
		{journal} {\bibinfo  {journal} {Bell System Technical Journal}\ }\textbf
		{\bibinfo {volume} {27}},\ \bibinfo {pages} {623} (\bibinfo {year}
		{1948})}\BibitemShut {NoStop}%
	\bibitem [{\citenamefont {Kolmogorov}(1968)}]{Kolmogorov1968}%
	\BibitemOpen
	\bibfield  {author} {\bibinfo {author} {\bibfnamefont {A.~N.}\ \bibnamefont
			{Kolmogorov}},\ }\href {\doibase 10.1080/00207166808803030} {\bibfield
		{journal} {\bibinfo  {journal} {International Journal of Computer
				Mathematics}\ }\textbf {\bibinfo {volume} {2}},\ \bibinfo {pages} {157}
		(\bibinfo {year} {1968})}\BibitemShut {NoStop}%
	\bibitem [{\citenamefont {Downarowicz}(2011)}]{downarowicz2011entropy}%
	\BibitemOpen
	\bibfield  {author} {\bibinfo {author} {\bibfnamefont {T.}~\bibnamefont
			{Downarowicz}},\ }\href@noop {} {\emph {\bibinfo {title} {{Entropy in
					dynamical systems}}}},\ Vol.~\bibinfo {volume} {18}\ (\bibinfo  {publisher}
	{Cambridge University Press},\ \bibinfo {year} {2011})\BibitemShut {NoStop}%
	\bibitem [{\citenamefont {Krieger}(1970)}]{krieger1970entropy}%
	\BibitemOpen
	\bibfield  {author} {\bibinfo {author} {\bibfnamefont {W.}~\bibnamefont
			{Krieger}},\ }\href@noop {} {\bibfield  {journal} {\bibinfo  {journal}
			{Transactions of the American Mathematical Society}\ }\textbf {\bibinfo
			{volume} {149}},\ \bibinfo {pages} {453} (\bibinfo {year}
		{1970})}\BibitemShut {NoStop}%
	\bibitem [{\citenamefont {Henriques}\ \emph {et~al.}(2013)\citenamefont
		{Henriques}, \citenamefont {Gon{\c{c}}alves}, \citenamefont {Antunes},
		\citenamefont {Matias}, \citenamefont {Bernardes},\ and\ \citenamefont
		{Costa-Santos}}]{Henriques2013}%
	\BibitemOpen
	\bibfield  {author} {\bibinfo {author} {\bibfnamefont {T.}~\bibnamefont
			{Henriques}}, \bibinfo {author} {\bibfnamefont {H.}~\bibnamefont
			{Gon{\c{c}}alves}}, \bibinfo {author} {\bibfnamefont {L.}~\bibnamefont
			{Antunes}}, \bibinfo {author} {\bibfnamefont {M.}~\bibnamefont {Matias}},
		\bibinfo {author} {\bibfnamefont {J.}~\bibnamefont {Bernardes}}, \ and\
		\bibinfo {author} {\bibfnamefont {C.}~\bibnamefont {Costa-Santos}},\ }\href
	{\doibase 10.1111/jep.12068} {\bibfield  {journal} {\bibinfo  {journal}
			{Journal of Evaluation in Clinical Practice}\ }\textbf {\bibinfo {volume}
			{19}},\ \bibinfo {pages} {1101} (\bibinfo {year} {2013})}\BibitemShut
	{NoStop}%
	\bibitem [{\citenamefont {Aboy}\ \emph {et~al.}(2006)\citenamefont {Aboy},
		\citenamefont {Hornero}, \citenamefont {Ab{\'{a}}solo},\ and\ \citenamefont
		{{\'{A}}lvarez}}]{Aboy2006}%
	\BibitemOpen
	\bibfield  {author} {\bibinfo {author} {\bibfnamefont {M.}~\bibnamefont
			{Aboy}}, \bibinfo {author} {\bibfnamefont {R.}~\bibnamefont {Hornero}},
		\bibinfo {author} {\bibfnamefont {D.}~\bibnamefont {Ab{\'{a}}solo}}, \ and\
		\bibinfo {author} {\bibfnamefont {D.}~\bibnamefont {{\'{A}}lvarez}},\ }\href
	{\doibase 10.1109/TBME.2006.883696} {\bibfield  {journal} {\bibinfo
			{journal} {IEEE Transactions on Biomedical Engineering}\ }\textbf {\bibinfo
			{volume} {53}},\ \bibinfo {pages} {2282} (\bibinfo {year}
		{2006})}\BibitemShut {NoStop}%
	\bibitem [{\citenamefont {Benedetto}\ \emph {et~al.}(2002)\citenamefont
		{Benedetto}, \citenamefont {Caglioti}, \citenamefont {Loreto},\ and\
		\citenamefont {Loreto}}]{Benedetto2002}%
	\BibitemOpen
	\bibfield  {author} {\bibinfo {author} {\bibfnamefont {D.}~\bibnamefont
			{Benedetto}}, \bibinfo {author} {\bibfnamefont {E.}~\bibnamefont {Caglioti}},
		\bibinfo {author} {\bibfnamefont {V.}~\bibnamefont {Loreto}}, \ and\ \bibinfo
		{author} {\bibfnamefont {V.}~\bibnamefont {Loreto}},\ }\href {\doibase
		10.1103/PhysRevLett.88.048702} {\bibfield  {journal} {\bibinfo  {journal}
			{Physical Review Letters}\ }\textbf {\bibinfo {volume} {88}},\ \bibinfo
		{pages} {4} (\bibinfo {year} {2002})}\BibitemShut {NoStop}%
	\bibitem [{\citenamefont {Amig{\'{o}}}\ \emph {et~al.}(2004)\citenamefont
		{Amig{\'{o}}}, \citenamefont {Szczepa{\'{n}}ski}, \citenamefont {Wajnryb},\
		and\ \citenamefont {Sanchez-Vives}}]{Amigo2004}%
	\BibitemOpen
	\bibfield  {author} {\bibinfo {author} {\bibfnamefont {J.~M.}\ \bibnamefont
			{Amig{\'{o}}}}, \bibinfo {author} {\bibfnamefont {J.}~\bibnamefont
			{Szczepa{\'{n}}ski}}, \bibinfo {author} {\bibfnamefont {E.}~\bibnamefont
			{Wajnryb}}, \ and\ \bibinfo {author} {\bibfnamefont {M.~V.}\ \bibnamefont
			{Sanchez-Vives}},\ }\href {\doibase 10.1162/089976604322860677} {\bibfield
		{journal} {\bibinfo  {journal} {Neural Computation}\ }\textbf {\bibinfo
			{volume} {16}},\ \bibinfo {pages} {717} (\bibinfo {year} {2004})}\BibitemShut
	{NoStop}%
	\bibitem [{\citenamefont {Melchert}\ and\ \citenamefont
		{Hartmann}(2015)}]{Melchert2015}%
	\BibitemOpen
	\bibfield  {author} {\bibinfo {author} {\bibfnamefont {O.}~\bibnamefont
			{Melchert}}\ and\ \bibinfo {author} {\bibfnamefont {A.~K.}\ \bibnamefont
			{Hartmann}},\ }\href {\doibase 10.1103/PhysRevE.91.023306} {\bibfield
		{journal} {\bibinfo  {journal} {Physical Review E}\ }\textbf {\bibinfo
			{volume} {91}},\ \bibinfo {pages} {023306} (\bibinfo {year}
		{2015})}\BibitemShut {NoStop}%
	\bibitem [{\citenamefont {Vogel}\ \emph {et~al.}(2012)\citenamefont {Vogel},
		\citenamefont {Saravia},\ and\ \citenamefont {Cortez}}]{Vogel2012}%
	\BibitemOpen
	\bibfield  {author} {\bibinfo {author} {\bibfnamefont {E.}~\bibnamefont
			{Vogel}}, \bibinfo {author} {\bibfnamefont {G.}~\bibnamefont {Saravia}}, \
		and\ \bibinfo {author} {\bibfnamefont {L.}~\bibnamefont {Cortez}},\ }\href
	{\doibase 10.1016/j.physa.2011.09.005} {\bibfield  {journal} {\bibinfo
			{journal} {Physica A: Statistical Mechanics and its Applications}\ }\textbf
		{\bibinfo {volume} {391}},\ \bibinfo {pages} {1591} (\bibinfo {year}
		{2012})}\BibitemShut {NoStop}%
	\bibitem [{\citenamefont {Vogel}\ \emph {et~al.}(2017)\citenamefont {Vogel},
		\citenamefont {Saravia},\ and\ \citenamefont {Ramirez-Pastor}}]{Vogel2017}%
	\BibitemOpen
	\bibfield  {author} {\bibinfo {author} {\bibfnamefont {E.~E.}\ \bibnamefont
			{Vogel}}, \bibinfo {author} {\bibfnamefont {G.}~\bibnamefont {Saravia}}, \
		and\ \bibinfo {author} {\bibfnamefont {A.~J.}\ \bibnamefont
			{Ramirez-Pastor}},\ }\href {\doibase 10.1103/PhysRevE.96.062133} {\bibfield
		{journal} {\bibinfo  {journal} {Physical Review E}\ }\textbf {\bibinfo
			{volume} {96}},\ \bibinfo {pages} {062133} (\bibinfo {year}
		{2017})}\BibitemShut {NoStop}%
	\bibitem [{\citenamefont {Martiniani}\ \emph {et~al.}(2019)\citenamefont
		{Martiniani}, \citenamefont {Chaikin},\ and\ \citenamefont
		{Levine}}]{Martiniani2019}%
	\BibitemOpen
	\bibfield  {author} {\bibinfo {author} {\bibfnamefont {S.}~\bibnamefont
			{Martiniani}}, \bibinfo {author} {\bibfnamefont {P.~M.}\ \bibnamefont
			{Chaikin}}, \ and\ \bibinfo {author} {\bibfnamefont {D.}~\bibnamefont
			{Levine}},\ }\href {\doibase 10.1103/PhysRevX.9.011031} {\bibfield  {journal}
		{\bibinfo  {journal} {Phys. Rev. X}\ }\textbf {\bibinfo {volume} {9}},\
		\bibinfo {pages} {011031} (\bibinfo {year} {2019})}\BibitemShut {NoStop}%
	\bibitem [{sup()}]{supp}%
	\BibitemOpen
	\href@noop {} {\emph {\bibinfo {title} {See Supplemental Material, which includes Refs.
				[33-36]}}}\BibitemShut {NoStop}%
	\bibitem [{\citenamefont {Cover}(2006)}]{CoverTM2006EOIT}%
	\BibitemOpen
	\bibfield  {author} {\bibinfo {author} {\bibfnamefont {T.~M.}\ \bibnamefont
			{Cover}},\ }\href@noop {} {\emph {\bibinfo {title} {Elements of information
				theory 2nd ed.}}}\ (\bibinfo  {publisher} {Wiley-Interscience},\ \bibinfo
	{address} {Hoboken, N.J.},\ \bibinfo {year} {2006})\BibitemShut {NoStop}%
	\bibitem [{\citenamefont {Ziv}\ and\ \citenamefont {Lempel}(1977)}]{Ziv1977}%
	\BibitemOpen
	\bibfield  {author} {\bibinfo {author} {\bibfnamefont {J.}~\bibnamefont
			{Ziv}}\ and\ \bibinfo {author} {\bibfnamefont {A.}~\bibnamefont {Lempel}},\
	}\href {\doibase 10.1109/TIT.1977.1055714} {\bibfield  {journal} {\bibinfo
			{journal} {IEEE Transactions on Information Theory}\ }\textbf {\bibinfo
			{volume} {23}},\ \bibinfo {pages} {337} (\bibinfo {year} {1977})}\BibitemShut
	{NoStop}%
	\bibitem [{\citenamefont {Ziv}\ and\ \citenamefont {Lempel}(1978)}]{Ziv1978}%
	\BibitemOpen
	\bibfield  {author} {\bibinfo {author} {\bibfnamefont {J.}~\bibnamefont
			{Ziv}}\ and\ \bibinfo {author} {\bibfnamefont {A.}~\bibnamefont {Lempel}},\
	}\href {\doibase 10.1109/TIT.1978.1055934} {\bibfield  {journal} {\bibinfo
			{journal} {IEEE Transactions on Information Theory}\ }\textbf {\bibinfo
			{volume} {24}},\ \bibinfo {pages} {530} (\bibinfo {year} {1978})}\BibitemShut
	{NoStop}%
	\bibitem [{\citenamefont {Pu}(2005)}]{pu2005fundamental}%
	\BibitemOpen
	\bibfield  {author} {\bibinfo {author} {\bibfnamefont {I.~M.}\ \bibnamefont
			{Pu}},\ }\href@noop {} {\emph {\bibinfo {title} {Fundamental data
				compression}}}\ (\bibinfo  {publisher} {Butterworth-Heinemann},\ \bibinfo
	{year} {2005})\ Chap.~\bibinfo {chapter} {7}\BibitemShut {NoStop}%
	\bibitem [{\citenamefont {Lesne}\ \emph {et~al.}(2009)\citenamefont {Lesne},
		\citenamefont {Blanc},\ and\ \citenamefont {Pezard}}]{Lesne2009}%
	\BibitemOpen
	\bibfield  {author} {\bibinfo {author} {\bibfnamefont {A.}~\bibnamefont
			{Lesne}}, \bibinfo {author} {\bibfnamefont {J.-L.}\ \bibnamefont {Blanc}}, \
		and\ \bibinfo {author} {\bibfnamefont {L.}~\bibnamefont {Pezard}},\ }\href
	{\doibase 10.1103/PhysRevE.79.046208} {\bibfield  {journal} {\bibinfo
			{journal} {Physical Review E}\ }\textbf {\bibinfo {volume} {79}},\ \bibinfo
		{pages} {046208} (\bibinfo {year} {2009})}\BibitemShut {NoStop}%
	\bibitem [{\citenamefont {Moon}\ \emph {et~al.}(2001)\citenamefont {Moon},
		\citenamefont {Jagadish}, \citenamefont {Faloutsos},\ and\ \citenamefont
		{Saltz}}]{Moon2001}%
	\BibitemOpen
	\bibfield  {author} {\bibinfo {author} {\bibfnamefont {B.}~\bibnamefont
			{Moon}}, \bibinfo {author} {\bibfnamefont {H.~V.}\ \bibnamefont {Jagadish}},
		\bibinfo {author} {\bibfnamefont {C.}~\bibnamefont {Faloutsos}}, \ and\
		\bibinfo {author} {\bibfnamefont {J.~H.}\ \bibnamefont {Saltz}},\ }in\ \href
	{\doibase 10.1109/69.908985} {\emph {\bibinfo {booktitle} {IEEE Transactions
				on Knowledge and Data Engineering}}},\ Vol.~\bibinfo {volume} {13}\ (\bibinfo
	{year} {2001})\ pp.\ \bibinfo {pages} {124--141}\BibitemShut {NoStop}%
	\bibitem [{\citenamefont {Plotnik}\ \emph {et~al.}(1992)\citenamefont
		{Plotnik}, \citenamefont {Weinberger},\ and\ \citenamefont
		{Ziv}}]{Plotnik1992UpperBounds}%
	\BibitemOpen
	\bibfield  {author} {\bibinfo {author} {\bibfnamefont {E.}~\bibnamefont
			{Plotnik}}, \bibinfo {author} {\bibfnamefont {M.~J.}\ \bibnamefont
			{Weinberger}}, \ and\ \bibinfo {author} {\bibfnamefont {J.}~\bibnamefont
			{Ziv}},\ }\href {\doibase 10.1109/18.108250} {\bibfield  {journal} {\bibinfo
			{journal} {IEEE Transactions on Information Theory}\ }\textbf {\bibinfo
			{volume} {38}},\ \bibinfo {pages} {66} (\bibinfo {year} {1992})}\BibitemShut
	{NoStop}%
	\bibitem [{\citenamefont {Singh}\ and\ \citenamefont
		{Warshel}(2010)}]{Singh2010}%
	\BibitemOpen
	\bibfield  {author} {\bibinfo {author} {\bibfnamefont {N.}~\bibnamefont
			{Singh}}\ and\ \bibinfo {author} {\bibfnamefont {A.}~\bibnamefont
			{Warshel}},\ }\href {\doibase 10.1002/prot.22689} {\bibfield  {journal}
		{\bibinfo  {journal} {Proteins: Structure, Function and Bioinformatics}\
		}\textbf {\bibinfo {volume} {78}},\ \bibinfo {pages} {1724} (\bibinfo {year}
		{2010})}\BibitemShut {NoStop}%
	\bibitem [{\citenamefont {Chiu}\ \emph {et~al.}(2005)\citenamefont {Chiu},
		\citenamefont {Kubelka}, \citenamefont {Herbst-Irmer}, \citenamefont {Eaton},
		\citenamefont {Hofrichter},\ and\ \citenamefont {Davies}}]{Chiu2005}%
	\BibitemOpen
	\bibfield  {author} {\bibinfo {author} {\bibfnamefont {T.~K.}\ \bibnamefont
			{Chiu}}, \bibinfo {author} {\bibfnamefont {J.}~\bibnamefont {Kubelka}},
		\bibinfo {author} {\bibfnamefont {R.}~\bibnamefont {Herbst-Irmer}}, \bibinfo
		{author} {\bibfnamefont {W.~a.}\ \bibnamefont {Eaton}}, \bibinfo {author}
		{\bibfnamefont {J.}~\bibnamefont {Hofrichter}}, \ and\ \bibinfo {author}
		{\bibfnamefont {D.~R.}\ \bibnamefont {Davies}},\ }\href {\doibase
		10.1073/pnas.0502495102} {\bibfield  {journal} {\bibinfo  {journal} {Proc.
				Natl. Acad. Sci. USA}\ }\textbf {\bibinfo {volume} {102}},\ \bibinfo {pages}
		{7517} (\bibinfo {year} {2005})}\BibitemShut {NoStop}%
	\bibitem [{\citenamefont {{Schr\"odinger, LLC}}(2015)}]{PyMOL}%
	\BibitemOpen
	\bibfield  {author} {\bibinfo {author} {\bibnamefont {{Schr\"odinger,
					LLC}}},\ }\href@noop {} {\enquote {\bibinfo {title} {The {PyMOL} molecular
				graphics system, version~2.1},}\ } (\bibinfo {year} {2015})\BibitemShut
	{NoStop}%
	\bibitem [{\citenamefont {Wolff}(1989)}]{Wolff1989}%
	\BibitemOpen
	\bibfield  {author} {\bibinfo {author} {\bibfnamefont {U.}~\bibnamefont
			{Wolff}},\ }\href {\doibase 10.1103/PhysRevLett.62.361} {\bibfield  {journal}
		{\bibinfo  {journal} {Physical Review Letters}\ }\textbf {\bibinfo {volume}
			{62}},\ \bibinfo {pages} {361} (\bibinfo {year} {1989})}\BibitemShut
	{NoStop}%
	\bibitem [{\citenamefont {Schlijper}\ and\ \citenamefont
		{Smit}(1989)}]{Schlijper1989}%
	\BibitemOpen
	\bibfield  {author} {\bibinfo {author} {\bibfnamefont {A.~G.}\ \bibnamefont
			{Schlijper}}\ and\ \bibinfo {author} {\bibfnamefont {B.}~\bibnamefont
			{Smit}},\ }\href {\doibase 10.1007/BF01044436} {\bibfield  {journal}
		{\bibinfo  {journal} {Journal of Statistical Physics}\ }\textbf {\bibinfo
			{volume} {56}},\ \bibinfo {pages} {247} (\bibinfo {year} {1989})}\BibitemShut
	{NoStop}%
	\bibitem [{\citenamefont {Onsager}(1944)}]{Onsager1944}%
	\BibitemOpen
	\bibfield  {author} {\bibinfo {author} {\bibfnamefont {L.}~\bibnamefont
			{Onsager}},\ }\href {\doibase 10.1103/PhysRev.65.117} {\bibfield  {journal}
		{\bibinfo  {journal} {Physical Review}\ }\textbf {\bibinfo {volume} {65}},\
		\bibinfo {pages} {117} (\bibinfo {year} {1944})}\BibitemShut {NoStop}%
	\bibitem [{\citenamefont {Wannier}(1950)}]{Wannier1950}%
	\BibitemOpen
	\bibfield  {author} {\bibinfo {author} {\bibfnamefont {G.~H.}\ \bibnamefont
			{Wannier}},\ }\href {\doibase 10.1103/PhysRev.79.357} {\bibfield  {journal}
		{\bibinfo  {journal} {Physical Review}\ }\textbf {\bibinfo {volume} {79}},\
		\bibinfo {pages} {357} (\bibinfo {year} {1950})}\BibitemShut {NoStop}%
\end{thebibliography}
\end{document}